\documentclass[11pt,a4paper,twoside]{article}
\usepackage[T1]{fontenc}
\usepackage[utf8]{inputenc}
\usepackage{amssymb}
\usepackage{lipsum}
\usepackage{authblk}
\usepackage[top=3cm, bottom=3cm, left=2cm, right=2cm]{geometry}
\usepackage{fancyhdr}
\usepackage{adjustbox}
\usepackage[sort&compress, numbers, compress ]{natbib}
\usepackage[colorlinks,bookmarksopen,bookmarksnumbered,citecolor=red,urlcolor=red]{hyperref}
\usepackage{algorithm}
\usepackage{algorithmic}
\usepackage{boxedminipage}
\usepackage{comment}
\usepackage{eqparbox}
\usepackage{multirow}
\usepackage{multicol}
\usepackage{caption}
\usepackage{subcaption}

\usepackage{nomencl}
\makenomenclature

\usepackage{amsmath}
\usepackage{amsfonts}

\usepackage{graphicx}

\usepackage{tikz}

\newlength{\commentindent}
\usepackage{array}
\newcolumntype{R}[1]{>{\raggedleft\let\newline\\\arraybackslash\hspace{0pt}}m{#1}}

\newcommand\BibTeX{{\rmfamily B\kern-.05em \textsc{i\kern-.025em b}\kern-.08em
T\kern-.1667em\lower.7ex\hbox{E}\kern-.125emX}}

\newcommand{\jump}[1]{[\![#1]\!]}
\newcommand{\avg}[1]{\{\!\!\{#1\}\!\!\}}

\newcommand{\feq}{f^{eq}}

\let\OLDthebibliography\thebibliography
\renewcommand\thebibliography[1]{
	\OLDthebibliography{#1}
	\setlength{\parskip}{0pt}
	\setlength{\itemsep}{0pt plus 0.3ex}
}

\renewenvironment{abstract}{%
	\begin{center}
		\begin{adjustbox}{minipage=0.90\textwidth,center} 
			\rule{\textwidth}{0.5pt}}
		{\par\noindent\rule{\textwidth}{0.5pt}  
		\end{adjustbox}
	\end{center} 
}

\makeatletter

\renewcommand\@maketitle{%
	\begin{adjustbox}{minipage=0.90\textwidth,center}
		\begin{center}
			\vskip 1em
			\let\footnote\thanks 
			{\LARGE \@title \par }
			\vskip 1.0em
			{\large \@author \par}
		\end{center}
	\end{adjustbox}
	\vskip 0.0em \par
}
\makeatother
\usepackage{etoolbox}
\renewcommand\nomgroup[1]{%
  \item[\bfseries
  \ifstrequal{#1}{G}{Greek Symbols}{%
  \ifstrequal{#1}{S}{Symbols}{%
  \ifstrequal{#1}{A}{Abbreviations}{%
  \ifstrequal{#1}{B}{Subscripts}{%
  \ifstrequal{#1}{P}{Superscripts}{
  }}}}}]}

\begin{document}

\title{Implicit Large Eddy Simulation of Nearly Incompressible Flows with a Discontinuous Galerkin-Boltzmann Formulation}
\author[1]{Onur~Ata\footnote{Corresponding author. E-mail address: onurata@metu.edu.tr} }
\author[1]{Atakan~Aygun}
\author[2]{Tim~Warburton}
\author[1]{Ali~Karakus}

\affil[1]{\textit{\small{Department of Mechanical Engineering, Middle East Technical University, Ankara, Turkey 06800}}}
\affil[2]{\textit{\small{Department of Mathematics, Virginia Tech, Blacksburg, VA, USA}}}

\renewcommand\Authands{ and }
\date{\vspace{-5ex}}
\maketitle

\begin{abstract} 
\textbf{Abstract}  
We present a high-order implicit large eddy simulation (ILES) approach for simulating flows at the nearly incompressible regime. Our methodology based on utilization of a nodal discontinuous Galerkin (DG) discretization of the Boltzmann equations. The compactness and low-dissipative nature of the discontinuous Galerkin method are leveraged to mimic traditional large eddy simulations with subgrid-scale models. One of the key requirements of ILES is to provide dissipation only within a narrow band of high wavenumbers. This is validated through numerical experiments on the Taylor-Green Vortex problem in detail at a Reynolds number where varying scales of coherent turbulent structures are present.
Furthermore, the approach is validated for external aerodynamic configurations by simulating the flow over a sphere at a Reynolds number of $Re=3700$, capturing the laminar–turbulent transition and the complex multiscale vortex dynamics characteristic of this regime. The results demonstrate the capability of the high-order DG-Boltzmann formulation to accurately capture transitional and turbulent flow features without the use of explicit sub-grid scale modeling, highlighting its potential as a robust and physically consistent framework for ILES of nearly incompressible turbulent flows.

\end{abstract}

% \input{nomen.tex}
% \printnomenclature

\section{Introduction}
Turbulent flows are inherently three-dimensional, unsteady, and random in nature. In these flows, significant and irregular variations in velocity scales occur both spatially and temporally \cite{pope2001turbulent}. Considering various parameters such as accuracy and computational cost, which generally depend on the specific flow problem, a range of numerical approaches has been developed to predict the behavior of turbulent flows. One such approach is the direct numerical simulation (DNS) \cite{orszag1972numerical,rogallo1981numerical}. This method resolves all scales of flow motion, from the largest scales—such as boundary layer or mixing layer thickness—down to the order of the Kolmogorov length scale, $\mathcal{O}(\eta)$, by solving the Navier-Stokes equations without using any turbulence models \cite{moin1998direct}. Although DNS is a powerful research tool for exploring turbulence physics, and significant advancements in computational power have been achieved in recent decades, it remains computationally prohibitive for most industrially relevant, complex flow problems.

Large eddy simulation (LES) has emerged as a viable alternative, among the various approaches for resolving the diverse scales of turbulent flow. While LES requires fewer computational resources than DNS, it provides greater insight into flow dynamics than the Reynolds-Averaged Navier-Stokes (RANS) methods that are more cost effective but less accurate. The traditional LES approach directly resolves the larger, three dimensional, unsteady motions that are anisotropic, geometry dependent and contain most of the energy. The smaller scales which have a more universal character are modeled  using subgrid-scale (SGS) models \cite{sagaut2009large}. As an alternative approach, numerical dissipation arising from the discretization schemes of numerical methods can be utilized to account for truncation errors that may represent SGS models. This approach is commonly referred to as implicit large eddy simulation (ILES) \cite{boris2005large, margolin2005design, margolin2006modeling} or also known as under-resolved DNS. Due to being a more direct approach by not requiring an SGS model, research interest in ILES has increased significantly over the last decade with various numerical methods \cite{moura2017eddy,wang2021implicit,matsuyama2023implicit}. 

A key requirement for ILES, and also all LES, is the ability to capture initial disturbances in the flow, which may be several orders of magnitude smaller than the freestream velocity. These disturbances cause laminar flows to transition into turbulence, leading to complex physical phenomena within the flow. Therefore, to accurately capture key structures in complex flows, a low level of dissipation concentrated within a narrow range at high wavenumbers is essential for ILES numerical methods \cite{drikakis2009large}. Thus, high-order methods are sensible choices for the accurate representation of turbulent flows because of their low-dissipative properties. Based on this, a broad class of discontinuous high order methods has been examined to assess their applicability to ILES through numerical experiments and analyses. These methods include discontinuous Galerkin (DG) \cite{hesthaven2007nodal,cockburn2012discontinuous}, spectral difference (SD) \cite{kopriva1996conservative,liu2006spectral,wang2007spectral,parsani2010implicit,zhou_implicit_2010}, spectral volume (SV) \cite{wang2002spectral}, and flux reconstruction (FR) \cite{huynh2007flux, bull2015simulation} methods. On the other hand, one of the main concern for suitability of these methods for ILES is their stability in terms of varying scenarios. In this context, these methods are analyzed and also compared with each other whether they provide adequately low dissipation while keep being numerically stable. 

Among high-order methods, flux reconstruction (FR) method has attracted attention due to its ability to operate with low numerical dissipation arising from its numerical flux formulation. This makes FR particularly suitable for under-resolved turbulence simulations. Studies have demonstrated that FR can capture key turbulence features even without explicit subgrid-scale models \cite{vermeire2016implicit}. Recent developments have further enabled its use in ILES setups through stabilization via additional filtering \cite{bull2015simulation, hamedi2022optimized}, anti-aliasing using approximate $L^2$ projection \cite{park2017high}, or hybrid RANS/LES formulations \cite{zhu2016implicit}. Our study focuses on DG method, which shares several structural and numerical properties with FR. The insights presented here may also be valuable for readers employing FR in ILES contexts.   

Discontinuous Galerkin (DG) methods are capable of producing high-order solutions for complex flows, a property that is particularly advantageous for large eddy simulations. Additionally, the compact nature of the high-order stencils used in the DG method facilitates the implementation of highly parallel algorithms on modern computing architectures efficiently. In the low Mach number regime, the application of the DG method to the compressible Navier–Stokes equations within the ILES framework has been extensively studied in the literature \cite{Uranga2011, carton2014assessment, beck2014high, bolemann2015high, Fernandez2017, winters2018comparative}. Also, it has been shown that robustness can be attained with various stabilization techniques \cite{ferrer2017interior, krank2017high, fehn2019high} in DG discretizations of the incompressible Navier–Stokes equations for ILES. Furthermore, the spectral properties of the DG method under various numerical flux formulations have been investigated in the context of ILES \cite{gassner2013accuracy, Moura2017, Fernandez2017, moura2017eddy, Mengaldo2018}, providing valuable insights and guiding further developments in this area. More recent studies continue to assess the performance of different DG methods for ILES and introduce new approaches aimed at achieving both accuracy and stability \cite{de2018use, fehn2019high, lederer2023high, moura2024joint}. In comparison to these recent advances in DG-based ILES for the Navier–Stokes equations, the analysis of DG discretizations of the Boltzmann equations within the ILES context remains unexplored. Several works have proposed DG formulations for the Boltzmann equation, demonstrating that numerical stability and accuracy can be maintained \cite{shi2003discontinuous, duster2006high, min2011spectral, Karakus2019}. Motivated by these results, we aim to explore the potential of using high-order DG-discretized Boltzmann equations to perform ILES for nearly incompressible flows.

In this paper, we investigate the capability of a high-order nodal DG discretization of the continuous Boltzmann equations \cite{Karakus2019} to perform accurate and stable implicit large-eddy simulations (ILES) of nearly incompressible three-dimensional turbulent flows. The primary objective is to assess how the intrinsic numerical dissipation and stabilization mechanisms of the method behave in under-resolved flow regimes. We argue that the relatively simple formulation of the Boltzmann equation system obtained with nodal DG method is well suited to control the dissipation mechanism and ensure stability. The governing equations are the Boltzmann equations with the Bhatnagar–Gross–Krook (BGK) collision operator \cite{bhatnagar1954model}. These equations are discretized in velocity space using a Galerkin approach with Hermite polynomials. The resulting first-order system, which includes a nonlinear collision operator, is solved using a nodal DG method. To reduce potential approximation errors in numerical integration and to ensure stabilization in coarse-grid simulations, the nonlinear collision operator is evaluated using a cubature rule as a de-aliasing mechanism. Numerical dissipation is provided exclusively by the discretization scheme at the appropriate temporal and spatial scales, removing the need for explicit filtering or artificial dissipation. Two different flux formulations ,upwind and local Lax-Friedrichs fluxes, are employed and compared to assess their effect on numerical dissipation for varying numbers of elements and approximation orders. The stability of the semi-discretized formulation is analyzed including the advective and nonlinear relaxation terms. A third-order semi-analytic Runge–Kutta scheme is used to alleviate the severe timestep restrictions that can arise in the presence of small relaxation times $\tau$. The capability of the proposed method to capture small-scale structures in transitional and turbulent flows is demonstrated using two benchmark problems: the Taylor–Green vortex at $Re=1600$ and flow around a sphere at $Re=3700$.

The remainder of this paper is organized as follows: In Section \ref{sec:Methodology}, we give details of the continuous Boltzmann equations and the discontinuous Galerkin discretization of the resulting system. We also discuss the details of the time discretization strategy. In Section \ref{sec:results}, we present the numerical studies to validate our approach and show the applicability of this system for large eddy simulations. Finally, we give some concluding remarks in Section \ref{sec:conclusion}.

\section{Methodology}
\label{sec:Methodology}
% In this section, we introduce the Galerkin-Boltzmann Formulation and give details about the discrete system. Then, we explain the high-order DG discretization and explicit time stepping methods.
In this section, first, we explain the basic concepts and the mathematics behind the Galerkin-Boltzmann formulation. We then go into the high-order discontinuous Galerkin methods focusing on the techniques for spatial discretization. Lastly, we cover the explicit time-stepping method used for time integration.
\subsection{Galerkin-Boltzmann Formulation}
We consider the Boltzmann equations with the BGK collision model. They formulate the time evaluation of a particle distribution function $f(\mathbf{x}, \mathbf{v}, t)$ which is a function of spatial variable $\mathbf{x}$, microscopic velocity $\mathbf{v}$, and time $t$. Neglecting external particle acceleration, the continuous Boltzmann-BGK equation can be written as

\begin{equation}
\label{eq:Boltzmann-BGK}
    \frac{\partial f}{\partial t} + \mathbf{v}\cdot\nabla_\mathbf{x} f =\frac{\left(f^{eq}-f\right)}{\tau},
\end{equation}
where $\tau$ is the relaxation time, which is a function of dynamic viscosity, and $\feq$ is the equilibrium phase space density written in three dimensions as
\begin{equation}
\label{eq:feq}
    \feq = \frac{\rho}{(2\pi RT)^{3/2}} \exp\left(-\frac{\left(\mathbf{v}-\mathbf{u}\right)^2}{2RT}\right),
\end{equation}
where $\rho, \; R,\;T$ and $\mathbf{u}$ are macroscopic density, gas constant, temperature, and macroscopic velocity, respectively.

To derive the Galerkin–Boltzmann equations, we start by adopting the approach of Tölke et al. \citep{tolke_discretization_2000}, approximating the phase space distribution function using a polynomial expansion based on the global steady-state flow.

\begin{equation}
    f(\mathbf{x}, \mathbf{v}, t) = \frac{1}{(2\pi RT)^{3/2}} \exp\left(-\frac{|\mathbf{v}|^2}{2RT} \right) \sum_{n=1}^{N_h}q_n(\mathbf{x},t)\phi_n(\mathbf{v}),
\end{equation}
where $\phi_n(\mathbf{v})$ are the polynomials in the velocity space and $q_n$ are the corresponding coefficients. Following a Galerkin formalism, we assume that $\phi_n$ functions form a basis, and the PDE residual is orthogonal to all test functions from the same polynomial space $\{ \phi_m(\mathbf{v})\}_{m=1}^{N_h}$. We multiply the equation with test functions and integrate in the unbounded microscopic velocity space $\Omega_{\mathbf{v}} = (-\infty,\infty)^d$ to obtain
\begin{equation}
    \label{eq:Galerkin_Boltzmann}
    \int_{\Omega_\mathbf{v}} \phi_m (\mathbf{v})\frac{1}{(2\pi R T)^{3/2}} \exp\left(-\frac{|\mathbf{v}|^2}{2RT}\right)\phi_n(\mathbf{v})\left( \frac{\partial q_n}{\partial t} +\mathbf{v}\cdot\nabla_\mathbf{x} q_n - \frac{\left(q^{eq}_n-q_n\right)}{\tau}\right)d\mathbf{v} = 0,
\end{equation}
where $\frac{1}{(2\pi R T)^{3/2}} \exp\left(-\frac{|\mathbf{v}|^2}{2RT}\right)\phi_n(\mathbf{v})q_n^{eq}$ is the local $L_2$ approximation of the local equilibrium state satisfying
\begin{equation}
\label{eq:Boltzmann-L2}
\int_{\Omega_\mathbf{v}} \phi_m(\mathbf{v})  \left( \frac{1}{(2\pi R T)^{3/2}} \exp\left(-\frac{|\mathbf{v}|^2}{2RT}\right)\phi_n(\mathbf{v}) q^{eq}_n  - f^{eq}\right)d\mathbf{v} = 0.
\end{equation}
We can write the equation system (\ref{eq:Galerkin_Boltzmann}) in the matrix form as 
\begin{equation}
 \label{eq:Boltzmann_matrix_form}
M_{mn}\frac{\partial q_n}{\partial t} = A_{\mathbf{x},mn}\cdot\nabla_\mathbf{x} q_n + M_{mn} \frac{1}{\tau} \left(q^{eq}_n - q_n\right),
\end{equation}
where the velocity space mass matrix, $M$, and stiffness matrices, $A_{\mathbf{x}}$, are given by
\begin{align}
 \label{eq:Boltzmann_5}
M_{mn} &=\frac{1}{(2\pi R T)^{3/2}} \int_{\Omega_\mathbf{v}} \phi_m \phi_n \exp\left(-\frac{|\mathbf{v}|^2}{2RT}\right)d\mathbf{v},\\ A_{\mathbf{x},mn} &= -\frac{1}{(2\pi R T)^{3/2}}\int_{\Omega_\mathbf{v}} \phi_m \phi_n \mathbf{v}\exp\left(-\frac{|\mathbf{v}|^2}{2RT}\right)d\mathbf{v}.
\end{align}
To simplify (\ref{eq:Boltzmann_matrix_form}), we choose Hermite polynomials to approximate the velocity space due to the exponential weighting in the inner products. Choosing tri-variate Hermite polynomials which are orthonormal under the Maxwellian inner product, leads to an identity mass matrix, $M_{mn}=\delta_{mn}$, and constant coefficient stiffness matrices. We then arrive at the following first order PDE
\begin{equation}
\label{eq:BoltzmannSystem}
\frac{\partial \mathbf{q}}{\partial t}= A_\mathbf{x}\cdot\nabla_\mathbf{x}\mathbf{q} + \mathcal{N}(\mathbf{q}),
\end{equation}
where $q = q(\mathbf{x},t)$ is a vector of Hermite polynomial coefficients, $A_\mathbf{x}$  are directional coefficient matrices, and $\mathcal{N}$ is the collision operator. We use second order Hermite polynomials to approximate the velocity space to model nearly incompressible flows in this paper. For a spatial dimension of three, second order velocity space approximation yields a vector of unknown coefficients of $\mathbf{q}(\mathbf{x},t) = [q_1(x,y,z,t), \dots, q_{10}(x,y,z,t)]$. The equations can be written explicitly as, 
% The stiffness matrices $A_{\mathbf{x}}=[A_x, A_y,A_z]$ and the nonlinear operator $\mathcal{N}$ becomes,

\begin{equation}
\setlength{\jot}{1.2ex} % space between rows
\begin{alignedat}{2}
\frac{\partial q_1}{\partial t} &= -c\left(\frac{\partial q_2}{\partial x} + \frac{\partial q_3}{\partial y} + \frac{\partial q_4}{\partial z}\right) 
&\quad \frac{\partial q_2}{\partial t} &= -c\left(\frac{\partial q_1}{\partial x} + \sqrt{2}\frac{\partial q_8}{\partial x} + \frac{\partial q_5}{\partial y} + \frac{\partial q_6}{\partial z}\right) \\[0.3em]
\frac{\partial q_3}{\partial t} &= -c\left(\frac{\partial q_5}{\partial x} + \frac{\partial q_1}{\partial y} + \sqrt{2}\frac{\partial q_9}{\partial y} + \frac{\partial q_7}{\partial z}\right) 
&\quad \frac{\partial q_4}{\partial t} &= -c\left(\frac{\partial q_6}{\partial x} + \frac{\partial q_7}{\partial y} + \frac{\partial q_1}{\partial z} + \sqrt{2}\frac{\partial q_{10}}{\partial z}\right) \\[0.3em]
\frac{\partial q_5}{\partial t} &= -c\left(\frac{\partial q_3}{\partial x} + \frac{\partial q_2}{\partial y} \right) -\frac{1}{\tau}\left(q_5 - \frac{q_2q_3}{q_1} \right) 
&\quad \frac{\partial q_6}{\partial t} &= -c\left(\frac{\partial q_4}{\partial x} + \frac{\partial q_2}{\partial z} \right)-\frac{1}{\tau}\left(q_6 - \frac{q_2q_4}{q_1} \right) \\[0.3em]
\frac{\partial q_7}{\partial t} &= -c\left(\frac{\partial q_4}{\partial y} + \frac{\partial q_3}{\partial z} \right) -\frac{1}{\tau}\left(q_7 - \frac{q_3q_4}{q_1} \right) 
&\quad \frac{\partial q_8}{\partial t} &= -c\left( \sqrt{2}\frac{\partial q_2}{\partial x}\right) - \frac{1}{\tau} \left(q_8 - \frac{q_2^2}{q_1\sqrt{2}} \right) \\[0.3em]
\frac{\partial q_9}{\partial t} &= -c\left( \sqrt{2}\frac{\partial q_3}{\partial y}\right) - \frac{1}{\tau} \left(q_9 - \frac{q_3^2}{q_1\sqrt{2}} \right) 
&\quad \frac{\partial q_{10}}{\partial t} &= -c\left( \sqrt{2}\frac{\partial q_4}{\partial z}\right) - \frac{1}{\tau} \left(q_{10} - \frac{q_4^2}{q_1\sqrt{2}} \right)
\end{alignedat}
\end{equation}
where $c=\sqrt{RT}$ represents the speed of sound. The system (\ref{eq:BoltzmannSystem}) recovers the Navier-Stokes equations for nearly incompressible flows with kinematic viscosity $\nu = \tau RT$ \cite{tolke_discretization_2000}. The macroscopic flow variables can be related to the microscopic variables as
%
% Return to default spacing
\setlength{\arraycolsep}{5pt}
\begin{equation*}
    \begin{matrix}
        \rho = q_1, & \rho u_x=\sqrt{RT}q_2, &\rho u_y=\sqrt{RT}q_3, & \rho u_z=\sqrt{RT}q_4
    \end{matrix}.
\end{equation*}
The components of the deviatoric stress tensor can be given as 
\begin{equation*}
    \begin{matrix}
        \sigma_{11} = -RT\left(\sqrt{2}q_8 - \frac{q_2^2}{q1} \right), &
        \sigma_{22} = -RT\left(\sqrt{2}q_9 - \frac{q_3^2}{q1} \right), & \sigma_{33} = -RT\left(\sqrt{2}q_{10} - \frac{q_4^2}{q1} \right), & \\
        \sigma_{12} = -RT\left(q_5 - \frac{q_2q_3}{q1} \right), & 
        \sigma_{13} = -RT\left(q_6 - \frac{q_2q_4}{q1} \right), & 
        \sigma_{23} = -RT\left(q_{7} - \frac{q_3q_4}{q1} \right),
    \end{matrix}.
\end{equation*}
The pressure is recovered with the equation of state $p=\rho RT$.

\subsection{Spatial Discretization}
We represent the physical domain $\Omega\in\mathbb{R}^3$ with computational domain $\Omega_h$ composed of $K$ non-overlapping elements, $D^e$, where $e=1,\dots,K$ such that
\begin{equation*}
    \Omega_h = \bigcup_{e=1}^KD^e.
\end{equation*}
We denote the approximation of field variable $q$ on element $D^e$ as $q^e$. The trace values of $q^e$ along $\partial D^e$ are denoted as $q^-$ and the corresponding neighboring values are denoted as $q^+$. We define the average and jump terms of $q^e$ along the trace $\partial D^e$ as $\avg{q}$ and $\jump{q}$ defined as 
\begin{equation}
    \begin{matrix}
        \avg{q} = \frac{q^+ + q^-}{2}, & \jump{q} = q^+ - q^-.
    \end{matrix}
\end{equation}
We consider a finite element space $V_N^e$ on each element $D^e$ to be $\mathcal{P}_N(D^e)$, the space of polynomial functions of degree $N$. As a basis of finite element spaces, we choose Lagrange polynomials to represent the solution on the grid points of each element, denoted as $\{\phi_i^e\}_{i=1}^{N_p}$. The Gauss-Legendre- (GLL) points are chosen as the interpolation nodes in the hexahedral elements. 

Next, starting from (\ref{eq:BoltzmannSystem}), we seek a local solution $q$ in an element $D^e$ satisfying
\begin{equation}
    \label{eq:variational_form}
    \int_{D^e}\phi\frac{\partial q}{\partial t} = \int_{D^e}\phi \left( A_x\frac{\partial q}{\partial x} + A_y\frac{\partial q}{\partial y} + A_z\frac{\partial q}{\partial z}\right) + \int_{\partial D^e} \phi F(q^*-q^-) +\int_{D^e}\phi\mathcal{N}(q).
\end{equation}
To evaluate the volume integral containing the nonlinear term $\mathcal{N}(q)$ in (\ref{eq:variational_form}), we use a cubature based integration to reduce aliasing errors. The cubature rule approximates the general integral for an arbitrary function $g$ in an element as
\begin{equation*}
    \int_{D^e} g(\mathbf{r}) \approx \sum_{i=1}^{N_c}g(\mathbf{r_i^c})w_i^c,
\end{equation*}
in $N_c$ cubature nodes with associated weights $w_i^c$ for $i=1,\dots,N_c$. We interpolate the microscopic coefficient field to a set of Gauss-Legendre cubature nodes to compute the nonlinear relaxation term $\mathcal{N}$ with the cubature weights $w_i^c$. 
% since it is shown that the spurious modes are expected to be negligible for upwind flux. In contrast, local Lax-Friedrichs flux can allow the existence of spurious reflections, especially for low Mach numbers \citep{Mengaldo2018}.

We selected the upwind flux as the numerical flux in our computations of the surface term. We denote the flux matrix $F = n_x A_x+n_y A_y +n_zA_z$ in the direction of the element normal vector $\mathbf{n}$ and $q^{*}$ is a trace state in equation (\ref{eq:variational_form}). The upwind flux can be formulated by diagonalizing the operator $F$ as $F =  \mathcal{R} \Lambda \mathcal{R}^{-1}$. The diagonal matrix, $\Lambda$ has only real entries of $0,0,0,0\pm c, \pm c, \pm c\sqrt{3}$. If we split the eigenvalues with positive signs as $\Lambda^+$ and negative signs as $\Lambda^-$, ie. $\Lambda=\Lambda^++\Lambda^-$ we can write the upwind flux as   

\begin{equation*}
    \label{eq:split_flux}
    F q^* = \mathcal{R}\left( \Lambda^{+} \mathcal{R}^{-1} q^- + \Lambda^{-} \mathcal{R}^{-1} q^+ \right).
\end{equation*}
We also compared the upwind flux scheme with the local Lax-Friedrichs (LLF) method to understand the effect of numerical flux formulation on the dissipative behavior of our methodology. The LLF intercell flux is given by
\begin{equation*}
    \label{eq:LLF_flux}
    F q^* = \frac{1}{2}\left(F^{+}+F^{-} \right)+\frac{1}{2}C \left(q^- - q^+ \right),
\end{equation*}
where $C$ is the local characteristic wave speed, and $F^{+}$ and $F^{-}$ are the corresponding components of the normal flux matrix constructed using the split eigenvalues, i.e., $F^+ = \mathcal{R} \Lambda^{+} \mathcal{R}^{-1}$ and $F^- = \mathcal{R} \Lambda^{-} \mathcal{R}^{-1}$. Unless otherwise stated, all results presented are obtained using the upwind flux scheme. 

By defining mass, surface mass, and stiffness operators as

\begin{equation*}
    \mathcal{M}^e_{ij}=\int_{D^e} \phi_j  \phi_i, \quad \mathcal{M}^{ef}_{ij}=\int_{\partial D^{e}} \phi_j  \phi_i, 
\end{equation*}

\begin{equation*}
(\mathcal{S}^e_{x})_{ij}=\int_{D^e} \phi_j \frac{\partial \phi_i}{\partial x} , \quad 
(\mathcal{S}^e_{y})_{ij}=\int_{D^e} \phi_j \frac{\partial \phi_i}{\partial y},
\quad
(\mathcal{S}^e_{z})_{ij}=\int_{D^e} \phi_j \frac{\partial \phi_i}{\partial z}, 
\end{equation*}
we write the equation (\ref{eq:variational_form}) in the semi-discrete form

\begin{equation}
\label{eq:semi_discrete1}
\mathcal{M}^e_{ij} \frac{\partial q^{e}_j}{\partial t}=A_x(\mathcal{S}^e_x)_{ij}q^e_j+A_y(\mathcal{S}^e_y)_{ij}q^e_j+A_z(\mathcal{S}^e_z)_{ij}q^e_j+\mathcal{M}^{ef}_{ij}(F(q^*-q^-))_j+\mathcal{J}^e\mathcal{I}_{ki}^ew_k\mathcal{N}(\mathcal{I}^e_{kj}q^e_j),   
\end{equation}
where $i,j=1,...N_p$ and $f=1,...N_{fp}$. With the multiplication of each term in the equation (\ref{eq:variational_form}) by inverse mass matrix, we can define the resulted differentiation, lifting, and projection operators as

\begin{equation*}
    \mathcal{D}^e_x=(\mathcal{M}^e)^{-1}\mathcal{S}^e_x,
    \quad
    \mathcal{D}^e_y=(\mathcal{M}^e)^{-1}\mathcal{S}^e_y,
    \quad
    \mathcal{D}^e_z=(\mathcal{M}^e)^{-1}\mathcal{S}^e_z,
\end{equation*}

\begin{equation*}
    \mathcal{L}^{ef}=(\mathcal{M}^e)^{-1}\mathcal{M}^{ef},
    \quad
    \mathcal{P}^e=(\mathcal{M}^e)^{-1}(\mathcal{I}^e)^T diag(w)
\end{equation*}
where $diag(w)$ is a diagonal matrix contains weights $w_i$ for $i=1,...,N_c$, then we can write semi discrete form (\ref{eq:semi_discrete1}) as

\begin{equation}
\label{eq:semi_discrete2}
 \frac{\partial q_i^e}{\partial t}=A_x(\mathcal{D}^e_x)_{ij}q_j+A_y(\mathcal{D}^e_y)_{ij}q_j+A_z(\mathcal{D}^e_z)_{ij}q_j+\mathcal{L}^e_{ij}(F(q^*-q^-))_j+\mathcal{J}^e\mathcal{P}^e_{ik}\mathcal{N}(\mathcal{I}^e_{kj}q^e_j)   
.\end{equation}
Finally, we express the nodal DG semi discrete form on the reference element $\hat{e}$ by reference operators

\begin{equation*}
\begin{aligned}
&\mathcal{D}^e_x=r^e_x\mathcal{D}_r+s^e_x\mathcal{D}_s+t^e_x\mathcal{D}_t, \\
&\mathcal{D}^e_y=r^e_y\mathcal{D}_r+s^e_y\mathcal{D}_s+t^e_y\mathcal{D}_t, \\
&\mathcal{D}^e_z=r^e_z\mathcal{D}_r+s^e_z\mathcal{D}_s+t^e_z\mathcal{D}_t,   \\ 
&\mathcal{L}^{ef}=\frac{\mathcal{J}^{ef}}{\mathcal{J}^e}\mathcal{L}^f, 
\quad
\mathcal{I}^e=\mathcal{I}, \\ 
&\mathcal{P}^e=\frac{1}{\mathcal{J}^e}\mathcal{P}, \\    
\end{aligned}
\end{equation*}
where we used the chain rule to compute physical derivatives $D^e_x$, $D^e_y$, $D^e_z$ by evaluating local coordinate derivatives $r_x^e$, $r_y^e$, $r_z^e$ etc. With the Lagrange basis polynomials $\phi_i$, we can write the partial derivative matrices on the reference element as
\begin{equation*}
\begin{aligned}
    \mathcal{D}_r &= D^{1D} \otimes I \otimes I, \\
    \mathcal{D}_s &= I \otimes D^{1D} \otimes I, \\
    \mathcal{D}_t &= I \otimes I \otimes D^{1D}, \\
\end{aligned}
\end{equation*}
%\[
%(G^e)^{-1}= \begin{bmatrix}
%r_x&r_y&r_z\\s_x&s_y&s_z\\t_x&t_y&t_z 
%\end{bmatrix}
%.\]
%
where, $D^{1D}$ is the one dimensional differentiation matrix and defined as tensor products of one dimensional polynomials.  Using these reference operators we write (\ref{eq:semi_discrete2}) as

\begin{equation}
\label{eq:semi_discrete_final}
 \frac{\partial q_i^e}{\partial t}=A^e_r(\mathcal{D}_r)_{ij}q_j+A^e_s(\mathcal{D}_s)_{ij}q_j+A^e_t(\mathcal{D}_t)_{ij}q_j+\frac{J^{ef}}{J^e}\mathcal{L}_{ij}(F(q^*-q^-))_j+\mathcal{P}_{ik}\mathcal{N}(I_{kj}q^e_j),   
\end{equation}

where
\begin{equation*}
\begin{aligned}
  &A^e_r=r^e_xA_x+r^e_yA_y+r^e_zA_z, \\
  &A^e_s=s^e_xA_x+s^e_yA_y+s^e_zA_z, \\
  &A^e_t=t^e_xA_x+t^e_yA_y+t^e_zA_z,      
\end{aligned}
\end{equation*}
completes the final form of the semi discrete nodal DG formulation we used. In the next sections, we complete our methodology by performing a stability analysis for the semi-discrete form and time discretization of these forms using a semi analytic approach.

\subsection{Stability of the Semi-Discrete Formulation}
Starting from the semi-discrete form shown in equation (\ref{eq:semi_discrete1})
\begin{equation*}
\mathcal{M}^e \frac{\partial q^{e}}{\partial t}=A_x(\mathcal{S}^e_x)q^e+A_y(\mathcal{S}^e_y)q^e+A_z(\mathcal{S}^e_zq^e+\mathcal{M}^{ef}(F(q^*-q^-))+\mathbf{N}^{e},
\end{equation*}
where $\mathbf{N}^{e}$ stands for discretized nonlinear collision term on element $e$.
Multiplying the equation with $q^T$ to convert it to energy form
\begin{equation*}
\frac{dE^e}{dt}=\boldsymbol{\Phi}_{\mathcal{L}}^e+\boldsymbol{\Phi}_{\mathcal{N}}^e=q^T(A_x(\mathcal{S}^e_x)q^e+A_y(\mathcal{S}^e_y)q^e +A_z(\mathcal{S}^e_z)q^e+\mathcal{M}^{ef}(F(q^*-q^-))+\mathbf{N}^{e}),   
\end{equation*}
where $E^{e}=\frac{1}{2}q^T\mathcal{M}q$, and $\boldsymbol{\Phi}_{\mathcal{L}}^e$ and $\boldsymbol{\Phi}_{\mathcal{N}}^e$ are the energy contribution of advection and nonlinear relaxation terms, respectively. Focusing on the linear term first, by using the summation by parts identity $S_{\mathbf{x}}+S_{\mathbf{x}}^T=\sum(R^{ef})^Tw^{ef}n_{\mathbf{x}}R^{ef}$ and symmetry of $A_{\mathbf{x}}$ we can write the discrete volume contribution term as
\[q^TA_{\mathbf{x}}S_{\mathbf{x}}q=\frac{1}{2}q^TA_{\mathbf{x}}(S_{\mathbf{x}}+S_{\mathbf{x}}^T)q=\frac{1}{2}\sum_f (q^{-})^{T}w^{ef}Fq^-\]
where $q^-=R^{ef}q$. Then substituting it into the energy form of the term 
\begin{equation*}
\boldsymbol{\Phi}_{\mathcal{L}}^e=\sum_f \left(\frac{1}{2}(q^{-})^{T}w^{ef}Fq^-+(q^{-})^{T}w^{ef}F(q^*-q^-)\right),   
\end{equation*}
Using upwind flux formulation and substituting it into energy form of the linear term, we obtain the following form of the energy on the faces of an element

\begin{equation} 
\boldsymbol{\Phi}_{\mathcal{L}}^e=\sum_f \left(-\frac{1}{2}(q^{-})^{T}w^{ef}(\mathcal{R}\Lambda \mathcal{R}^{-1})q^-+(q^{-})^{T}w^{ef}\mathcal{R}\left( \Lambda^{+} \mathcal{R}^{-1} q^- + \Lambda^{-} \mathcal{R}^{-1} q^+ \right)\right).
\end{equation}
If we add the contribution of the neighbour faces and sum over all unique interior faces in the domain we can obtain the following form of the linear term as  
\begin{equation}
% \frac{d E}{d t}=\left(\sum -\frac{1}{2}\jump{q}^T w^{ef}R\jump{\Lambda}R^{-1}\jump{q}\right),   
\boldsymbol{\Phi}_{\mathcal{L}}=\sum_{Nf} \left( -\frac{1}{2}\jump{q}^T w^{ef}\mathcal{R}\jump{\Lambda}\mathcal{R}^{-1}\jump{q}\right) \leq0,   
\end{equation}
since $\jump{F}=\mathcal{R}\jump{\Lambda}\mathcal{R}
^{-1}$ is positive semidefinite similarly shown in previous studies \cite{hesthaven2007nodal,warburton2013low,kopriva2014energy,chan2017weight,chan2017penalty}. 
Next, we focus on the nonlinear collision term to complete the analysis. The contribution to the elemental energy rate from the relaxation term can be written as 
\begin{equation*}
    \boldsymbol{\Phi}_{\mathcal{N}}^e = (q^e)^T\mathcal{J}^e\mathcal{I}_{ki}^ew_k\mathcal{N}(\mathcal{I}^e_{kj}q^e_j) = \sum_k w_k J_k^e(q(\xi_k))^T\mathcal{N}(q(\xi_k)).
\end{equation*}
Since the first four equation does not have any relaxation term, we will omit these equations. Each component can be written in the form
\begin{equation*}
    q_i(\xi_k) \mathcal{N}_i(\xi_k) = -\frac{1}{\tau} q_i(\xi_k) (q_i(\xi_k) - q_i(\xi_k)^{\text{eq}})
= -\frac{1}{\tau}\left[(q_i(\xi_k) - q_i^{\text{eq}}(\xi_k))^2 + q_i^{\text{eq}}(\xi_k)(q_i(\xi_k) - q_i^{\text{eq}}(\xi_k))\right].
\end{equation*}
Hence, elemental relaxation contribution gives 
\begin{equation*}
    \boldsymbol{\Phi}_{\mathcal{N}}^e = -\frac{1}{\tau}\sum_k\sum_{i=5}^{10}w_kJ_k^e\left[ (q_i(\xi_k) - q_i^{\text{eq}}(\xi_k))^2 + q_i^{\text{eq}}(\xi_k)(q_i(\xi_k) - q_i^{\text{eq}}(\xi_k))\right].
\end{equation*}
The leading quadratic term here is strictly non-positive $-\frac{1}{\tau}\sum_{i=5}^{10}(q_i(\xi_k) - q_i^{\text{eq}}(\xi_k))^2 \le 0$. The second term does not have a fixed sign but it can be bounded using Young's inequality,
\begin{equation*}
    \big|q_i^{\text{eq}}(\xi_k)(q_i(\xi_k) - q_i^{\text{eq}}(\xi_k)) \big| \le \frac{\varepsilon}{2}(q_i(\xi_k) - q_i^{\text{eq}}(\xi_k))^2 + \frac{1}{2\varepsilon}(q_i^{eq}(\xi_k))^2, \quad \varepsilon>0.
\end{equation*}
Substituting this bound to the energy form gives 
\begin{equation*}
    \boldsymbol{\Phi}_{\mathcal{N}}^e \le -\frac{1-\varepsilon/2}{\tau}\sum_k\sum_{i=5}^{10} w_kJ_k^e(q_i(\xi_k) - q_i^{\text{eq}}(\xi_k))^2 + \frac{1}{2\tau \varepsilon}\sum_k\sum_{i=5}^{10}w_kJ_k^e(q_i^{\text{eq}}(\xi_k))^2.
\end{equation*}
By summing over all elements, we can write the the total relaxation contribution to the discrete energy rate in the form
\begin{equation}
    \boldsymbol{\Phi}_{\mathcal{N}} \le -\frac{1-\varepsilon/2}{\tau}|| q - q^{\text{eq}}||^2_{L^2} + \frac{C}{\tau \varepsilon},
\end{equation}
where 
\begin{equation*}
    || q - q^{\text{eq}}||^2_{L^2} = \sum_e\sum_k\sum_{i=5}^{10}w_kJ_k^e(q_i(\xi_k) - q_i^{\text{eq}}(\xi_k))^2, \quad C = \frac{1}{2} \sum_e\sum_k\sum_{i=5}^{10}w_kJ_k^e(q_i^{\text{eq}}(\xi_k))^2.
\end{equation*}
Hence, for any physically admissible state (\(q_1>0\) and \(0<\varepsilon<2\)), the relaxation operator is dissipative up to a bounded source term proportional to \(\|q_i^{\rm eq}\|^2\). However, numerical quadrature errors can introduce spurious energy and lead to nonphysical solution growth, particularly in under-resolved simulations. To mitigate aliasing errors, we employ over-integration by evaluating the nonlinear terms using a higher-order quadrature rule. Combination of the analysis of the linear and nonlinear terms shows the stability of the semi-discrete form of the Galerkin-Boltzmann formulation.

\subsection{Temporal Discretization}
In our simulations, we have used third order semi analytic Runge-Kutta time discretization \citep{Karakus2019} to overcome the severe time step restriction in the limit of small relaxation time $\tau$. The nonlinear term $\mathcal{N}(q)$ in the formulation becomes stiff for small relaxation times, therefore, we write the system as
\begin{equation}
    \label{eq:time_split_stiffness}
    \frac{d\mathbf{q}}{dt} = \mathbf{L}(\mathbf{q}) + \mathbf{N}(\mathbf{q}),
\end{equation}
collecting all linear (non-stiff) terms in $\mathbf{L}$ and nonlinear (stiff) relaxation terms that depends on $\tau$ in $\mathbf{N}$. From the nonlinear term in (\ref{eq:BoltzmannSystem}), we can rewrite (\ref{eq:time_split_stiffness}) as 
\begin{equation}
    \frac{d\mathbf{q}}{dt} = -\boldsymbol{\Lambda}\mathbf{q} + \mathbf{L}(\mathbf{q}) + \Tilde{\mathbf{N}}(\mathbf{q}),
\end{equation}
where $\boldsymbol{\Lambda} = \mathrm{diag(0, 0, 0, 0, \frac{1}{\tau}, \frac{1}{\tau}, \frac{1}{\tau}, \frac{1}{\tau}, \frac{1}{\tau}, \frac{1}{\tau})}$ and $\Tilde{\mathbf{N}}(\mathbf{q}) = \left( 0, 0, 0, 0,\frac{q_2q_3}{\tau q_1}, \frac{q_2q_4}{\tau q_1},\frac{q_3q_4}{\tau q_1}, \frac{q_2^2}{\tau q_1\sqrt{2}}, \frac{q_3^2}{\tau q_1\sqrt{2}}, \frac{q_4^2}{\tau q_1\sqrt{2}}\right)^T$. By writing $\mathbf{F}(\mathbf{q}) = \mathbf{L}(\mathbf{q}) + \Tilde{\mathbf{N}}(\mathbf{q})$ we can simplify the notation as 
\begin{equation}
    \label{eq:time_split}
    \frac{d\mathbf{q}}{dt} = -\boldsymbol{\Lambda}\mathbf{q} + \mathbf{F}(\mathbf{q}).
\end{equation}
The structure of the system allows the first four equations to be integrated explicitly in time. This is feasible because the nonlinear term is absent, making the advective time scale the relevant stability constraint. The time-splitting approach in (\ref{eq:time_split}) separates the system such that we can apply a semi-analytic time integration method to the last six equations, while continuing to use an explicit method for the first four. We carefully design the explicit scheme so that, in the non-stiff limit $1/\tau\rightarrow0$, it aligns with the semi-analytic integration. To derive the semi-analytic formulation, we multiply (\ref{eq:time_split}) by $e^{\boldsymbol{\Lambda t}}$ and integrate in time, yielding the following expression:
\begin{equation}
    \label{eq:time_volterra}
    \mathbf{q}(t_{n+1})= \mathbf{q}(t_n) e^{-\boldsymbol\Lambda (t_{n+1}-t_n)} + \int_{t_n}^{t_{n+1}} e^{\boldsymbol\Lambda (\theta-t_{n+1}) }\mathbf{F}\left(\mathbf{q}\left(\theta\right), \theta \right)d\theta.
\end{equation}
Since the first four eigenvalues are zero, the first four equations of (\ref{eq:time_volterra}) are identical to the equations in (\ref{eq:time_split_stiffness}) integrated in time. For a Runge-Kutta method, we begin by integrating (\ref{eq:time_volterra}) from \( t = t_n \) to some intermediate time level \( t = t_n + \Delta t_i \),

\begin{equation*}
\mathbf{q}_{ni}= \mathbf{q}_n e^{-\boldsymbol\Lambda \Delta t_i} + \int_0^{\Delta t_i} e^{\boldsymbol\Lambda(\theta-\Delta t_i) }\mathbf{F}\left(\mathbf{q}( t_n+\theta) , t_n + \theta\right)d\theta.
\end{equation*}
Internal and final stages of a general method can be approximated by,
\begin{equation*}
\label{Eq:ERK_1}
    \begin{split}
    \mathbf{q}_{ni}&= \mathbf{q}_n e^{-\boldsymbol\Lambda \Delta t_i} + \Delta t\sum_{j=0}^{s-1}\tilde{a}_{ij}\mathbf{F}\left(\mathbf{q}( t_n+\Delta t_j) , t_n + \Delta t_j\right) = \mathbf{q}_n e^{-\boldsymbol\Lambda \Delta t_i} + \Delta t\sum_{j=0}^{s-1}\tilde{a}_{ij}\mathbf{F}_{nj}, \\
\mathbf{q}_{n+1}&= \mathbf{q}_n e^{-\boldsymbol\Lambda \Delta t} + \Delta t\sum_{i=0}^{s-1}\tilde{b}_{i}\mathbf{F}\left(\mathbf{q}( t_n+\Delta t_i) , t_n + \Delta t_i\right) = \mathbf{q}_n e^{-\boldsymbol\Lambda \Delta t} + \Delta t\sum_{j=0}^{s}\tilde{b}_{i}\mathbf{F}_{ni},
    \end{split}
\end{equation*}
\def\arraystretch{1.5}%  
\begin{table}[t]
\caption{Butcher tableau for the adapted method, based on RK2a, with coefficients }
    \setlength{\tabcolsep}{5pt}
      \centering
        \begin{tabular}{c| c c c}
            0             &                &  & \\
            $\frac{1}{3}$ & $\frac{1}{3}$  &  & \\
            $\frac{3}{4}$ & $-\frac{3}{16}$  & $\frac{15}{16}$ & \\ \hline
              & $\frac{1}{6}$ & $\frac{3}{10}$ &$\frac{8}{15}$
        \end{tabular}
    \label{table:Butcher}
\end{table}
where \( s \) is the number of stages, and \( \tilde{a} \) and \( \tilde{b} \) are the semi-analytic Runge-Kutta method coefficients. Since a semi-analytic method reduces to the base Runge-Kutta method in the limit \( \frac{1}{\tau} \rightarrow 0 \), the exponential and non-exponential terms are consistent in the equation. It is assumed that the base RK method satisfies
\begin{equation}
    \sum_{j=0}^{s-1} b_j = 1, \quad \sum_{j=0}^{s-1} a_{ij} = c_i,
\end{equation}
and the semi-analytic RK scheme satisfies an analogous constraint:
\begin{equation}
    \sum_{j=0}^{s-1} \tilde{b}_j = \gamma^{-1} \left( e^{\gamma} - 1 \right), \quad \sum_{j=0}^{s-1} \frac{1}{c_i} \gamma^{-1} \left( e^{c_i \gamma} - 1 \right),
\end{equation}
for \( i = 1, \dots, s-1 \) and \( \gamma = -\frac{\Delta t}{\tau} \). A class of third-order semi-analytic RK schemes was introduced in \cite{Karakus2019} by modifying the base method coefficients, resulting in improved truncation errors. With the Butcher tableau given in Table~\ref{table:Butcher}, the coefficients used in this work are given as,
\begin{equation}
    \label{eq:ERK_5}
    \begin{split}
        \tilde{a}_{10} &= \gamma^{-1} \left[-1 + e^{\frac{\gamma}{3}}\right], \\
        \tilde{a}_{20} &= \frac{1}{4}\gamma^{-1} \left[ 1 - e^{\frac{3\gamma}{4}}\right], \\
        \tilde{a}_{21} &= \frac{1}{4}\gamma^{-1} \left[-5 + 5e^{\frac{3\gamma}{4}}\right], \\
        \tilde{b}_{0}  &= \frac{1}{3}\gamma^{-3}\left[ -24 - 11\gamma - 2\gamma^2 - e^{\gamma}\left(-24 + 13\gamma - 3\gamma^2 \right) \right],\\
        \tilde{b}_{1}  &= \frac{36}{5}\gamma^{-3}\left[ 2 + \frac{5}{4}\gamma + \frac{1}{4}\gamma^2 - e^{\gamma}\left(2 - \frac{3}{4}\gamma\right) \right],\\
        \tilde{b}_{2}  &= \frac{16}{5}\gamma^{-3}\left[ -2 - \frac{5}{3}\gamma - \frac{2}{3}\gamma^2 - e^{\gamma}\left(-2 + \frac{1}{3}\gamma\right) \right],
    \end{split}
\end{equation} 
% 
% In the limit of $\frac{1}{\tau}\rightarrow 0$, this method reduces to the base Runge-Kutta method which makes the integrated equation consistent.

\section{Results}
\label{sec:results}
The purpose of this section is to validate the accuracy and stability of the Boltzmann-Galerkin methodology for different types of flows. In this context, two different cases are examined: the three-dimensional Taylor-Green Vortex (TGV) problem and the flow over a sphere at $Re = 3700$.
% transitional flow around the Selig-Donovan 7003 airfoil, and turbulent flow around a non-rotating golf ball. 
All calculations were performed within the libParanumal framework \cite{ChalmersKarakusAustinSwirydowiczWarburton2020}, where the DG Boltzmann formulation is implemented with the semi-analytic Runge-Kutta scheme.

 \begin{figure}[htbp!]
     \centering
         \begin{subfigure}[b]{0.48\textwidth}
             \includegraphics[width=\textwidth]{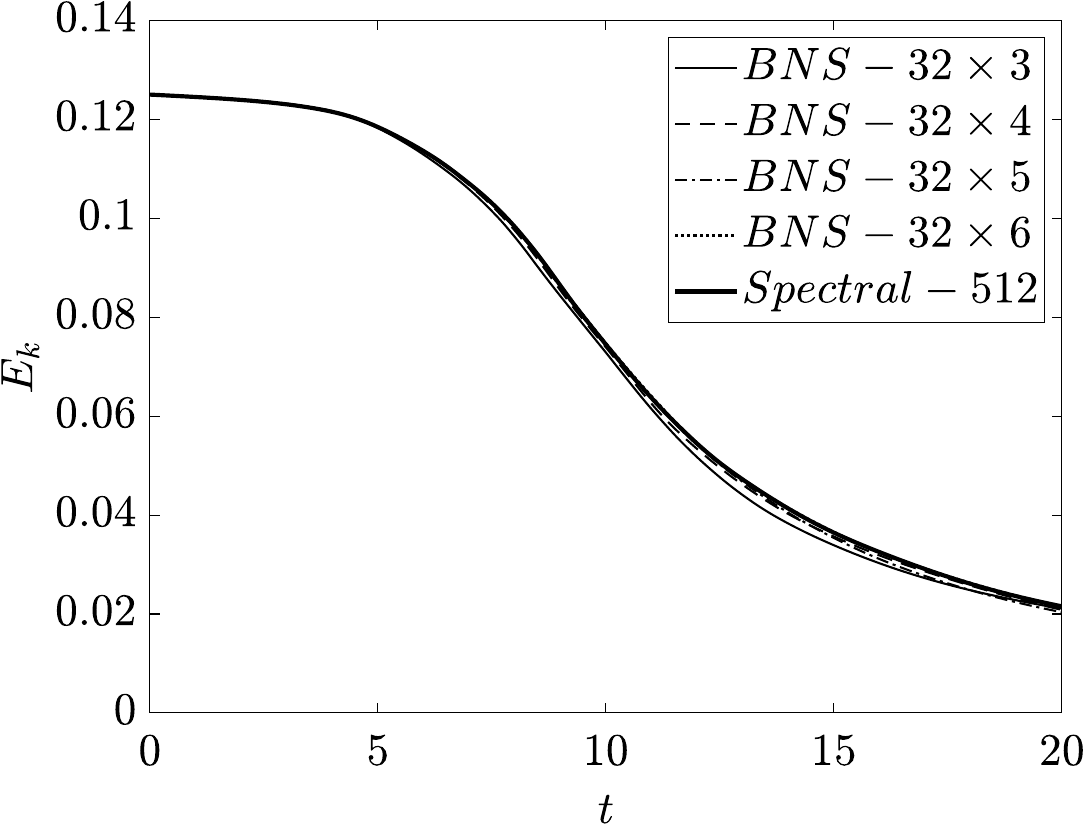}    
         \end{subfigure}
         \begin{subfigure}[b]{0.48\textwidth}
             \includegraphics[width=\textwidth]{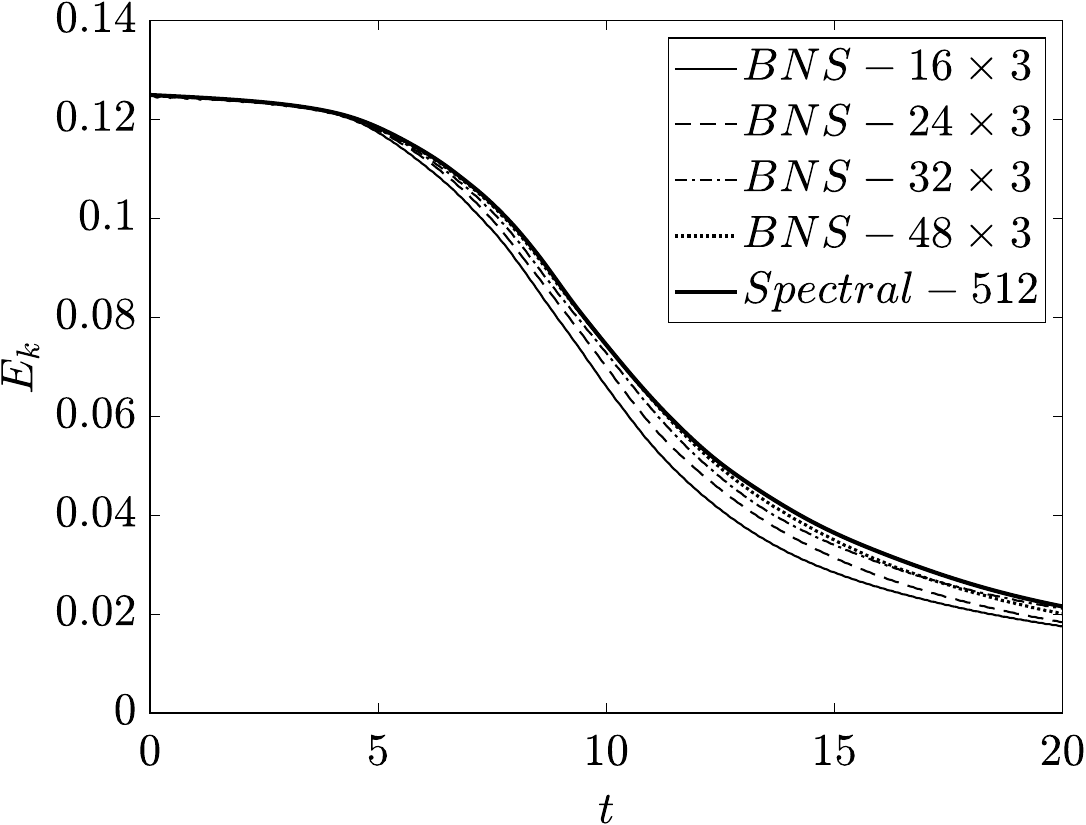}
         \end{subfigure}
     \caption{Evolution of the kinetic energy for 3D TGV problem. Left: comparison of different orders of approximation. Right: comparison of different numbers of elements.}
     \label{fig:KE_order}
 \end{figure}

\subsection{3D Taylor-Green Vortex} 
The Taylor-Green Vortex is a problem to study the relation between dissipation of energy and decay of isotropic turbulence \cite{taylor1937mechanism}. It is a commonly used benchmark to assess the numerical method's capability of solving turbulent flows since it has simple initial and boundary conditions and presents key aspects of a complex turbulent flow. The solution domain in this problem is a periodic cubic box with dimensions $x_{i}\in[0,2\pi]$. Reynolds number for this case is selected as $1600$. Our methodology solves nearly incompressible flows. Thus, Mach number is selected as $0.1$. Initial conditions of the problem are
\begin{align}
    u(x,y,z,t_0)&=U_{0}sin(x/L)cos(y/L)cos(z/L), \\  
    v(x,y,z,t_0)&=-U_{0}sin(y/L)cos(x/L)cos(z/L), \\
    w(x,y,z,t_0)&=0.
\end{align}
The solution domain is discretized using equispaced hexahedral elements. In this context, four different meshes are used, consisting of elements $16^{3}$, $24^{3}$, $32^{3}$, and $48^{3}$. We obtain solutions for increasing number of elements at a moderate order, $N=3$, and increasing orders for number of elements $K=32^3$. The results are compared with those obtained using a spectral element method with $512^{3}$ elements \cite{van2011comparison}. In the plots, the reference solution is denoted as $Spectral-512$. All the solutions for TGV problem is obtained on Nvidia P100 GPUs paired with Intel Xeon Gold 6248R processors on the TUBITAK ULAKBIM, High Performance and Grid Computing Center.

\begin{figure}[htbp!]
    \centering
        \begin{subfigure}[b]{0.48\textwidth}
            \includegraphics[width=\textwidth]{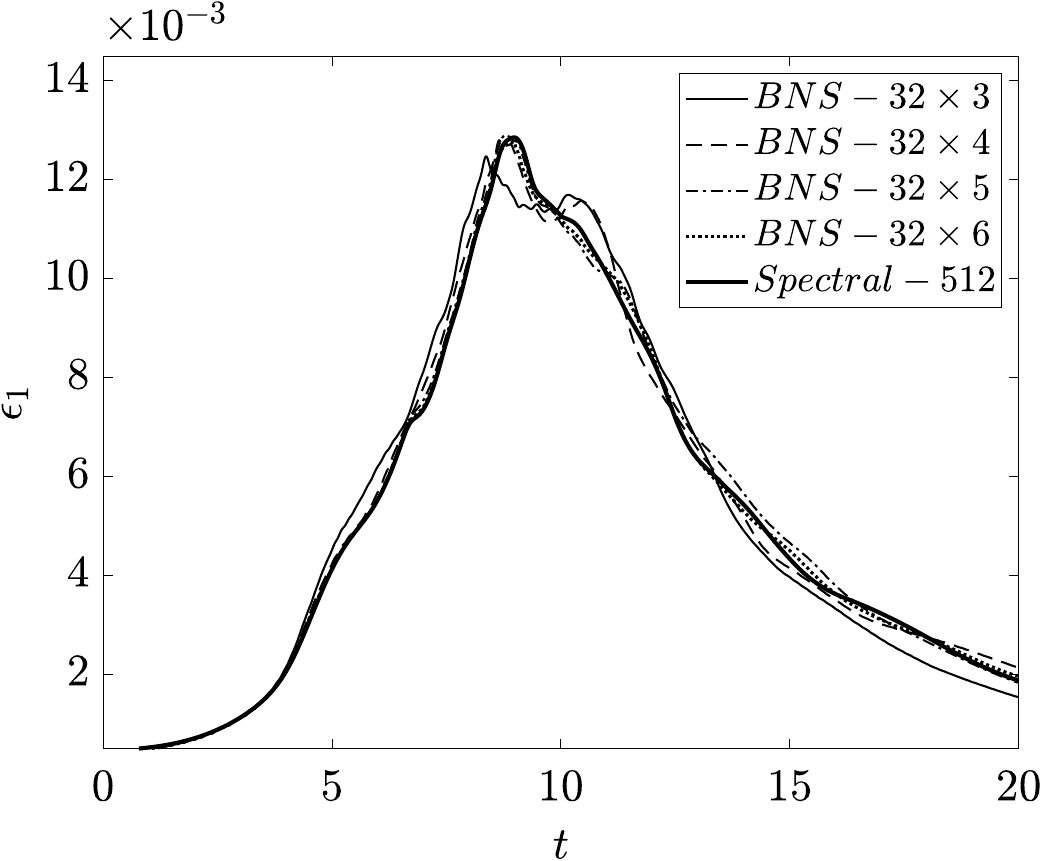}
        \end{subfigure}
        \begin{subfigure}[b]{0.48\textwidth}
            \includegraphics[width=\textwidth]{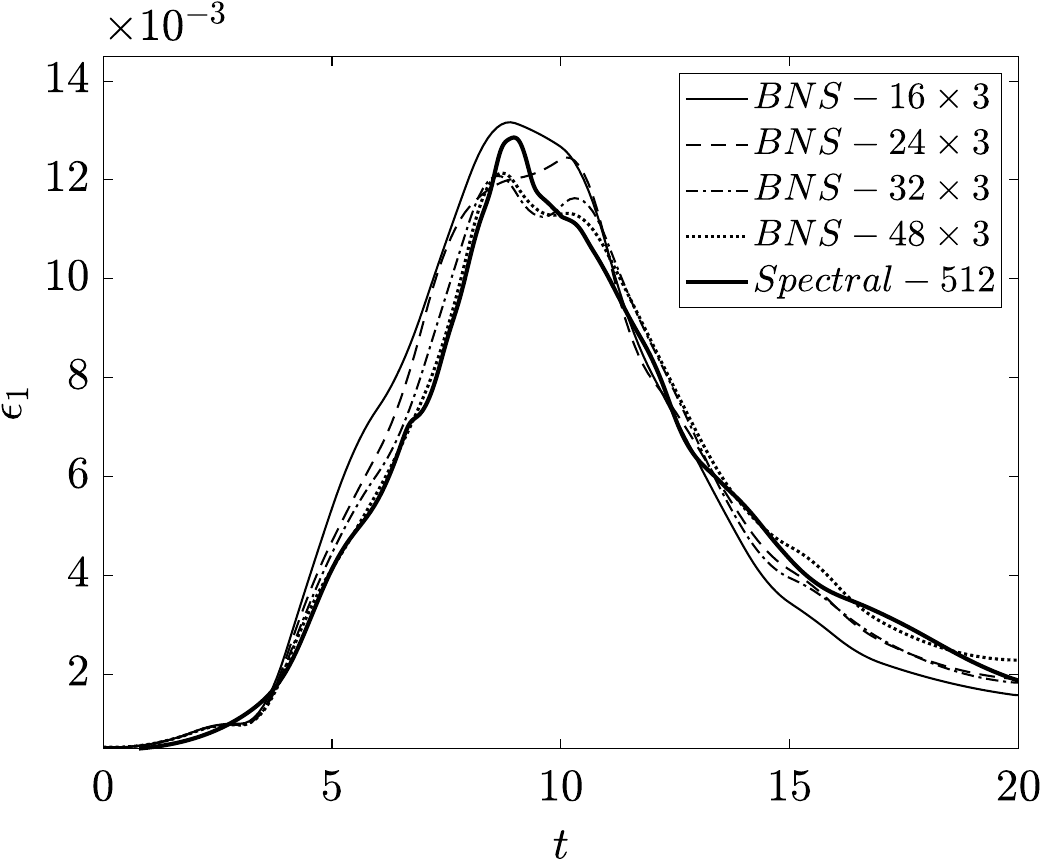}    
        \end{subfigure}
    \caption{Kinetic energy dissipation rate for 3D TGV problem. Comparison of different orders of approximation.}
    \label{fig:KEDR_order}
\end{figure}

The TGV problem is analyzed to evaluate how the kinetic energy in the domain dissipates over time. The total kinetic energy in the domain is computed as
\begin{equation}
    E_{k}=\frac{1}{\rho_{\infty}\Omega}\int_{\Omega}\frac{1}{2}\rho \mathbf{u\cdot u}d\Omega,
\end{equation}
where $\Omega$ is the volume of the domain and the reference density $\rho_{\infty}=1$. The time evolution of the kinetic energy is shown in Figure \ref{fig:KE_order} between $0\le t\le 20$. The solutions with $K=16^3, \ 24^3, \ 32^3, \ 48^3$ and $N=3$ overestimate the energy dissipation through the smaller scales. Solutions with a moderate mesh, $K=32^3$ and increasing orders of approximation $N=3,4,5,6$ present less dissipation, and sixth order solution with $32^3$ elements well agreed with the reference solution.

\begin{figure}[htbp!]
    \centering
        \begin{subfigure}[b]{0.48\textwidth}
            \includegraphics[width=\textwidth]{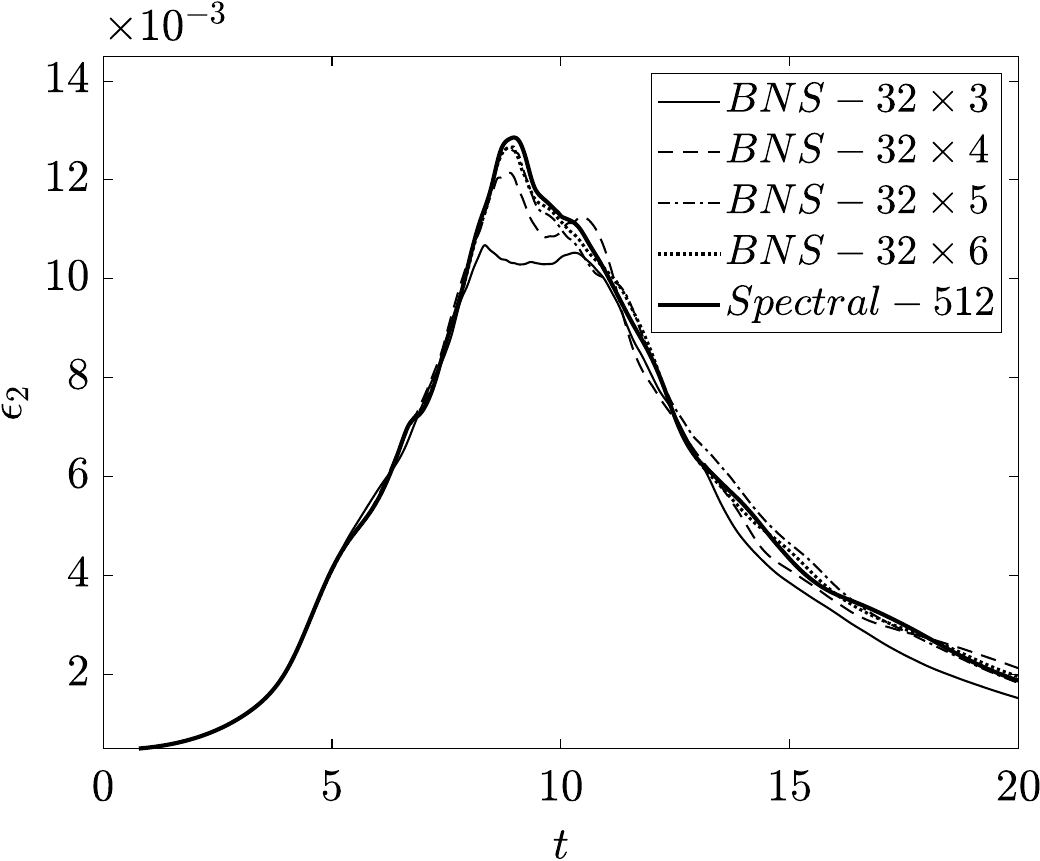}
        \end{subfigure}
        \begin{subfigure}[b]{0.48\textwidth}
            \includegraphics[width=\textwidth]{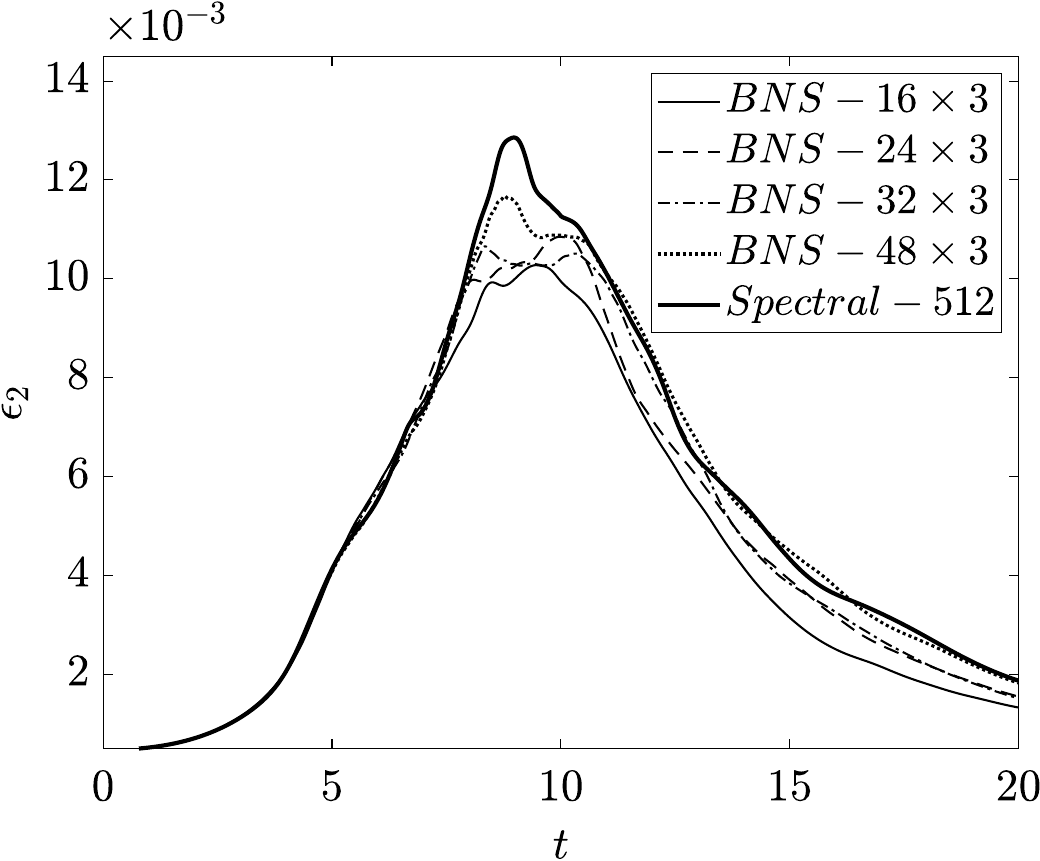}
        \end{subfigure}    
    \caption{Kinetic energy dissipation rate based on enstrophy for 3D TGV problem. Comparison of different orders of approximation.}
    \label{fig:Enstrophy_order}
\end{figure}

The kinetic energy dissipation reads as 
\begin{equation}
    \epsilon_{1}=-\frac{dE_{k}}{dt},
\end{equation}
where it is computed directly by a first order forward differencing of kinetic energy. The second dissipation term is related to enstrophy. Enstrophy can be found by
\begin{equation}
     \zeta=\frac{1}{\rho_{\infty}\Omega}\int_{\Omega}\frac{1}{2}\rho \boldsymbol{\omega \cdot \omega} d\Omega,
\end{equation}
where $\boldsymbol{\omega}$ is vorticity. For incompressible flows, a second dissipation rate can also be found by using enstrophy as
\begin{equation}
    \epsilon_{2}=-\frac{2\mu}{\rho_{\infty}}\zeta,
\end{equation}
where $\mu$ is the dynamic viscosity. Dissipation rates computed using both methods are shown in figures \ref{fig:KEDR_order} and \ref{fig:Enstrophy_order}. A discrepancy is observed between the two approaches: the vorticity-based dissipation rate is consistently lower than the dissipation rate calculated directly from kinetic energy. As suggested in the literature \cite{debonis2013solutions}, this difference indicates that dissipation originates not only from vorticity but also from numerical dissipation that is defined by $\epsilon_1-\epsilon_2$.

\begin{figure}[htbp!]
    \centering
    \includegraphics[width=0.48\textwidth]{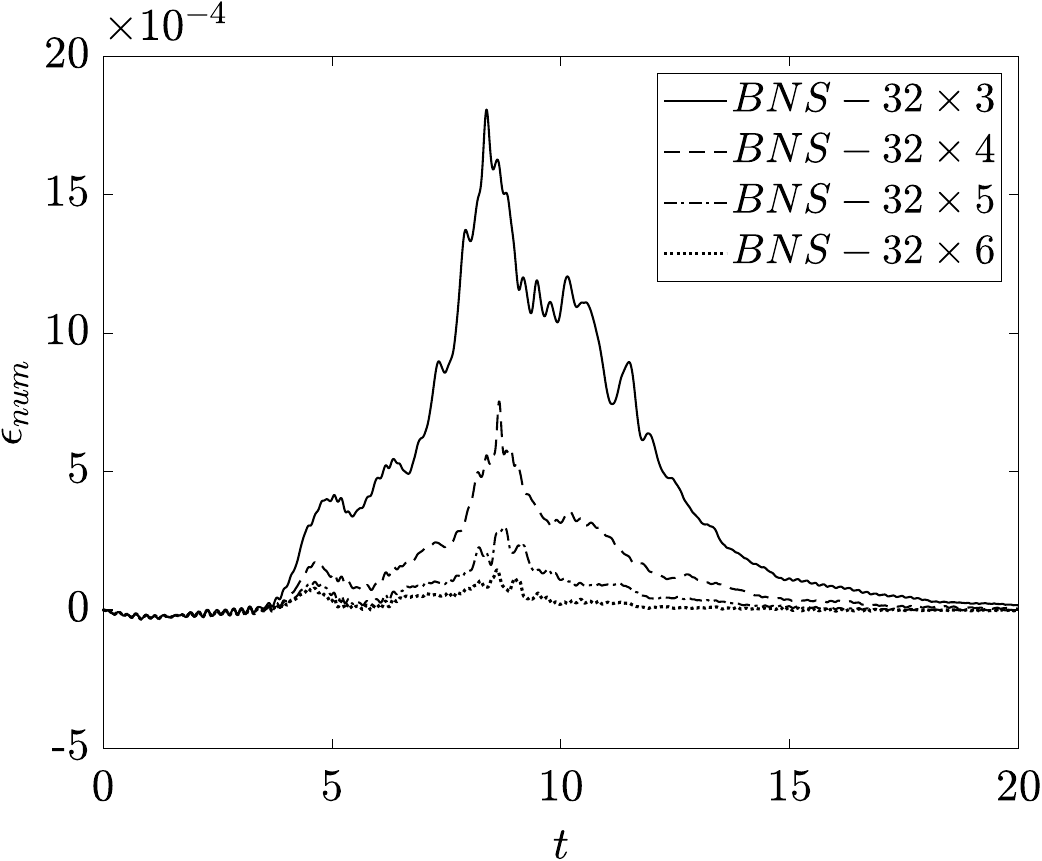}
    \caption{Numerical dissipation rate for increasing orders of approximation for 3D TGV Problem.}
    \label{fig:Dissipation_Difference}
\end{figure}

\begin{figure}[htbp!]
    \centering
        \begin{subfigure}[b]{0.45\textwidth}
            \includegraphics[width=\textwidth]{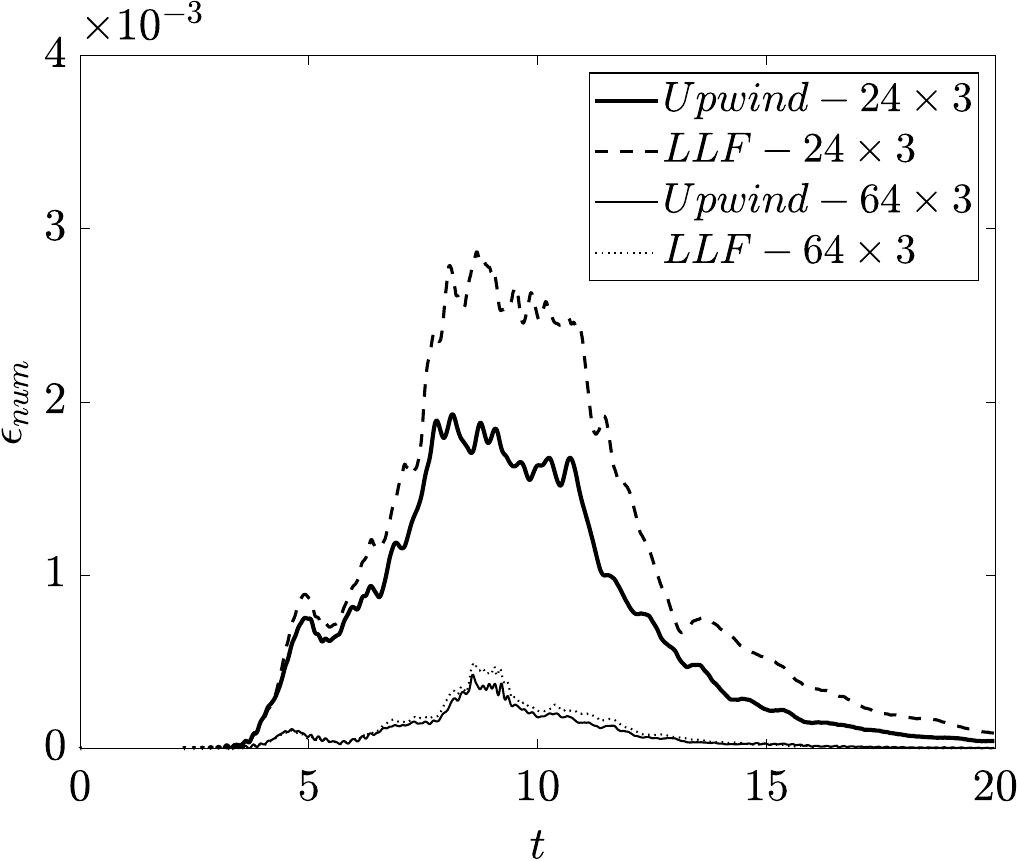}
        \end{subfigure}
        \begin{subfigure}[b]{0.45\textwidth}

            \includegraphics[width=\textwidth]{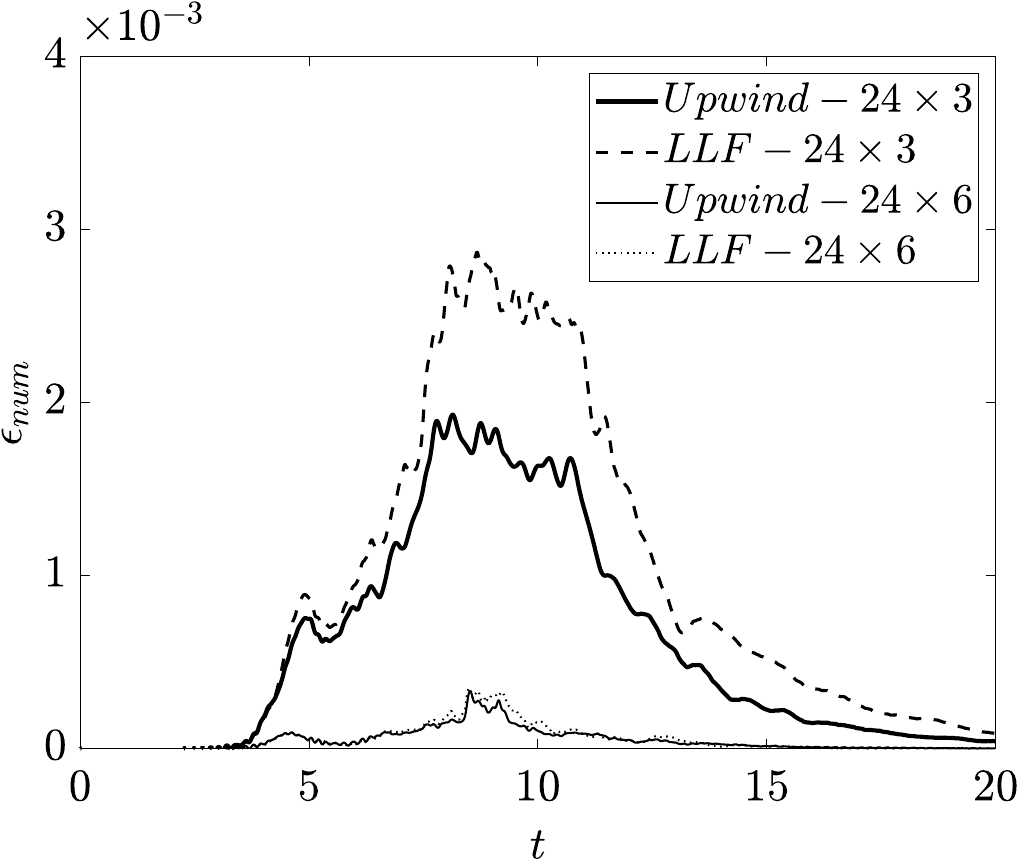}
        \end{subfigure}

\caption{Comparison of numerical dissipation rates for different numerical flux approaches with varying numbers of elements and orders of approximation for the 3D TGV problem. Left: for increasing number of elements, right: increasing orders of approximation.}

    \label{fig:Dissipation_Difference_flux}
\end{figure}

The solutions with $K=32^3$ and $N=4,5,6$, predict kinetic energy dissipation rate computed directly from kinetic energy accurately as it is shown in the Figure \ref{fig:KEDR_order}. Comparisons with the reference solution indicate that all these configurations successfully capture the peak dissipation around $t=9$. Increasing the number of elements improves the results to the moderate number of elements $K=32$, but further refinement shows less improvement on the accurate presentation of the dissipation rate. Further, the dissipation due to enstrophy shown in Figure \ref{fig:Enstrophy_order} increases as smaller vortical structures are better resolved with higher polynomial orders, and the solutions for $N=5$ and $N=6$ show strong agreement with the reference results. These findings suggest that the numerical dissipation provided by the scheme effectively mimics the behavior of an SGS model. A comparison of numerical dissipation for increasing polynomial orders is presented in Figure \ref{fig:Dissipation_Difference}. The peak numerical dissipation around mid-time is consistently observable as a general trend. For polynomial approximation orders $N=5$ and $N=6$, very low numerical dissipation is evident, enabling the generation of stable and accurate high-order ILES solutions. Numerical dissipation is used as a comparative quantity to observe the effect of the numerical flux formulation. In this context, the classic upwind scheme is compared with the LLF scheme. The relationship between the numerical flux and total numerical dissipation is analyzed using three different configurations. In the first configuration, both the number of elements and the polynomial order are relatively small, $K=24^3$, $N=3$. Under these conditions, the LLF scheme generates considerable numerical dissipation. In the other two configurations, the effects of increasing $K$ and $N$ are investigated. The results, also shown in Figure~\ref{fig:Dissipation_Difference_flux}, indicate that while increasing the number of elements still leads to some difference in numerical dissipation—especially at the peak point—increasing the order of approximation suppresses the excess dissipation introduced by the LLF scheme.

\begin{figure}[htbp!]
    \centering
        \begin{subfigure}[b]{0.35\textwidth}
            \includegraphics[width=\textwidth]{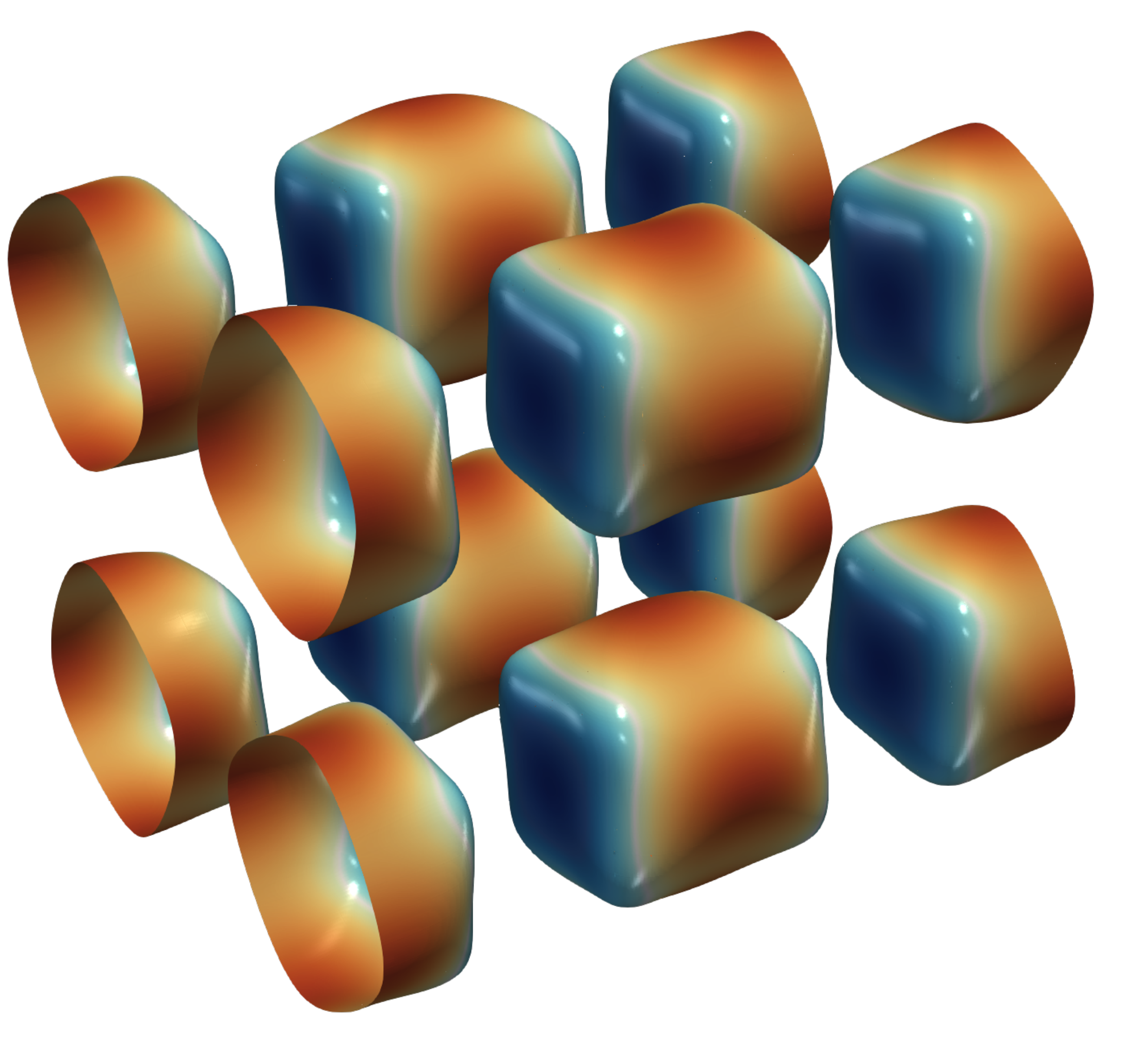}
        \end{subfigure}
        %\begin{subfigure}[b]{0.30\textwidth}
        %    \includegraphics[width=\textwidth]{figures/t=5.pdf}
        %\end{subfigure}
        \begin{subfigure}[b]{0.35\textwidth}
            \includegraphics[width=\textwidth]{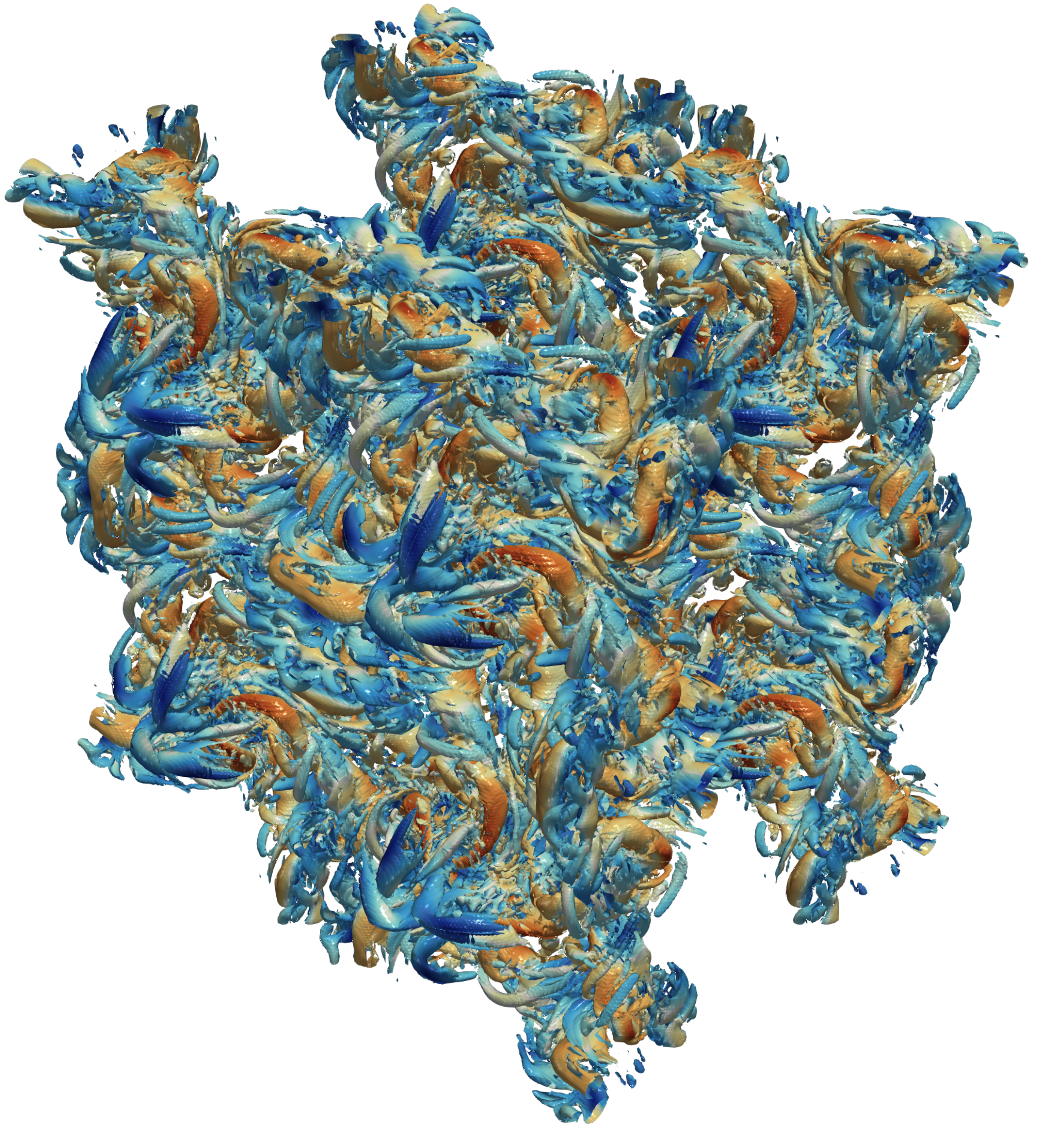}
        \end{subfigure}
\caption{Iso-surfaces of the Q-criterion, colored by velocity magnitude, from the $K=32^3$, $N=6$ solution of the 3D TGV problem. Left: $t^*=1$; right: $t^*=9$.}

    \label{fig:Q_Crit_TGV}
\end{figure}

 To visualize coherent structures in the flow, especially in the wake region,  Q-criterion \cite{hunt1988eddies} is used. It is calculated by the positive second invariant of the velocity gradient tensor \({\nabla} \mathbf{u} \), defined as
\[Q = \frac{1}{2}(\|\boldsymbol{\Omega}\|^2-\|\mathbf{S}\|^2).\]
where \(\mathbf{S}\) and \(\boldsymbol{\Omega}\) are symmetric and antisymmetric parts of velocity gradient \({\nabla} \mathbf{u} \), respectively. Iso-contours of the Q-criterion, colored by velocity magnitude, are shown in Figure~\ref{fig:Q_Crit_TGV} to illustrate the flow evolution. As described by \cite{brachet1983small}, the initial inviscid behavior develops into the formation and roll-up of vortex sheets up to $t^* = 5$. Coherent vortex structures then begin to break down around $t^* = 9$. Beyond this point, the flow becomes fully turbulent, and the structures decay until the flow reaches a steady state.

\begin{figure}[htbp!]
    \centering
        \begin{subfigure}[b]{0.48\textwidth}
            \includegraphics[width=\textwidth]{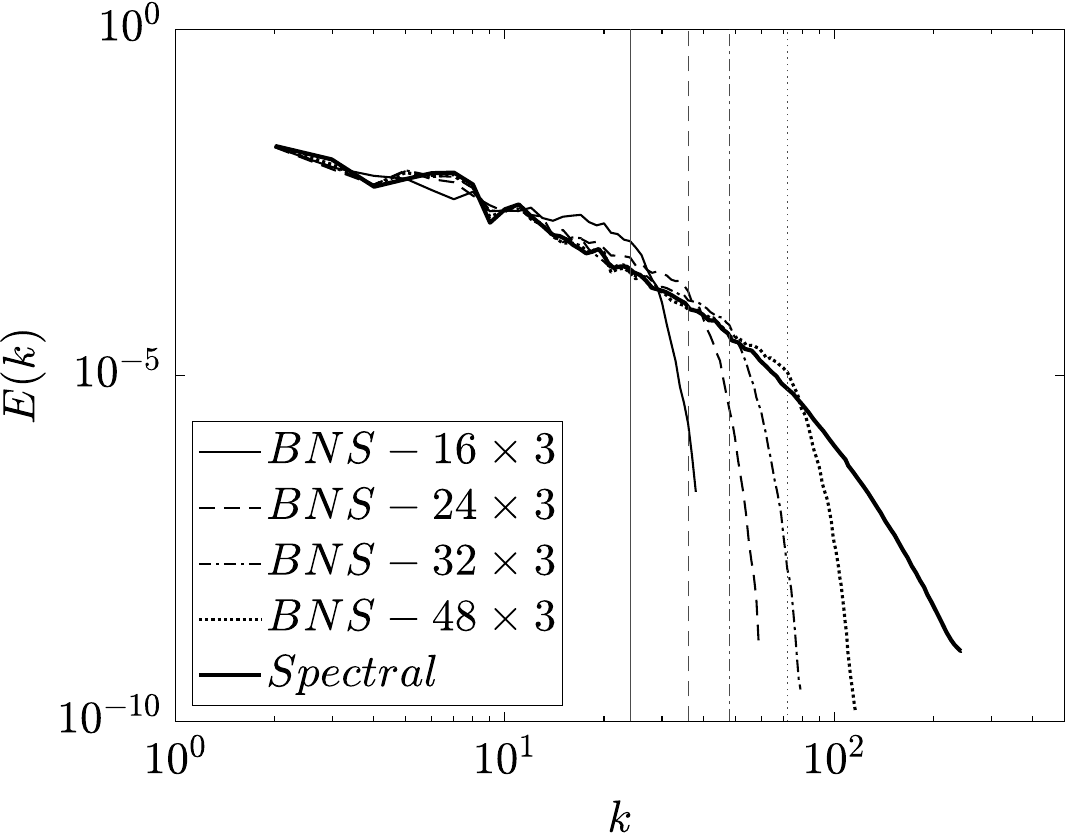}
        \end{subfigure}
        \begin{subfigure}[b]{0.48\textwidth}
            \includegraphics[width=\textwidth]{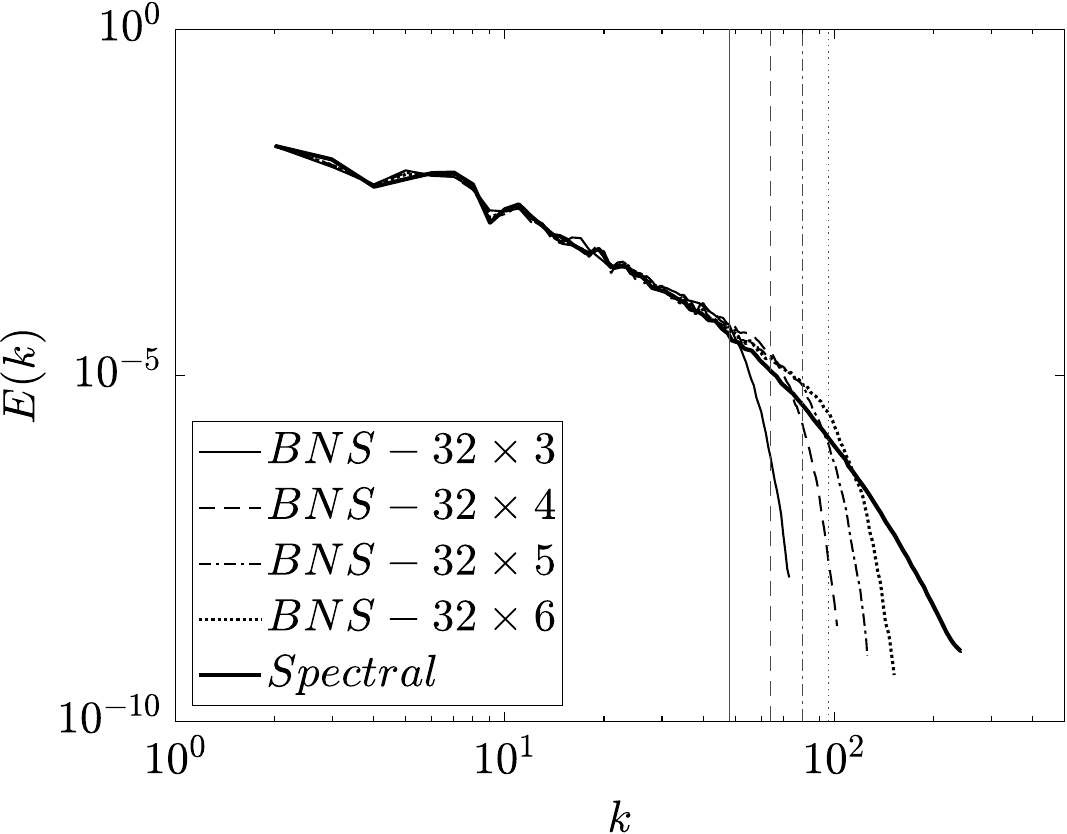}
        \end{subfigure}
    \caption{Energy Spectra at t=9 for 3D TGV problem.}
    \label{fig:Energy Spectrum}
\end{figure}

The three-dimensional kinetic energy spectra is also calculated using the Fourier transform as a function of the wavenumber. Face values are averaged to represent the spectra accurately. Figure \ref{fig:Energy Spectrum} shows the energy spectra for solutions with varying numbers of elements and polynomial orders at $t=9$ where the energy in the smallest scales peaks and maximum dissipation appears around that time \cite{brachet1983small}. The results are compared with a reference solution obtained using a pseudospectral method with $512^{3}$ elements \cite{carton2014assessment}. The maximum wavenumber for each simulation, given by $DOF/2$, is indicated by vertical lines in the plots. A comparison of an increasing number of elements and polynomial orders suggests that using higher-order polynomials with a moderate number of elements is favorable for capturing energy at higher wavenumbers. Furthermore, for the coarsest mesh, $K=16^3$, noticeable inconsistencies are observed at lower wavenumbers as it is also observed in \cite{gassner2013accuracy}.                                          

\subsection{Flow over a sphere at \(Re=3700\)}

In this section, we investigate the flow over a sphere in the sub-critical regime at \(Re=3700\). We compare our results with previous numerical studies \cite{yun2006vortical, rodriguez2011direct, zhang2022dynamics} and experimental data \cite{kim1988observations,sakamoto1990study} for the same regime. At this Reynolds number, the flow separates laminarly near the equator of the sphere and becomes turbulent in the separated shear layer, resulting in vortex shedding. The transition from laminar to turbulent flow, together with the unsteady vortex shedding in the turbulent wake, makes the flow dynamics complex. 

The computational domain is discretized using hexahedral elements. To preserve solution accuracy at the wall boundary, curved elements are employed on the sphere surface. These curved elements are obtained by relocating the surface nodes to coincide with the sphere boundary, while the interior volume nodes are adjusted by applying a simple blending method. In total, $K=476199$ elements are used to obtain the solution with third-order and fourth-order polynomial approximations. We used perfectly matched layers (PML) to damp out the reflections from the boundaries and keep the computational domain relatively small. The solution is obtained with a PML width $w=4D$ in all directions, maximum damping coefficient of $50$ and a third order damping profile, where $D$ is the sphere diameter. For more detailed information and formulation of PML, readers can refer \cite{Karakus2019}. The simulations were run on NVIDIA A100 GPUs paired with AMD EPYC 7713 processors on the Polaris supercomputer. 

The mean flow parameters, drag coefficient, Strouhal number, and instantaneous vortical structures are investigated and compared with reference data. The drag and lift coefficients, \(C_d\) and \(C_l\), are obtained from the total force, $\mathbf{F}_t$, acting on the sphere surface, \(\Gamma\),
\[
\mathbf{F}_t=\int_\Gamma (-\boldsymbol{\sigma} \cdot \mathbf{n} + p \mathbf{n}) \, d\Gamma,
\] 
where $p$ is the pressure recovered from the equation of state and \(\boldsymbol{\sigma}\) is the deviatoric stress tensor, with components defined in the methodology section. The coefficients are then defined as 
\[
C_d=\frac{\mathbf{F}_t \cdot \mathbf{i}}{\tfrac{1}{2}\rho_{\infty}U^2_{\infty}A}, \quad 
C_l=\frac{\mathbf{F}_t \cdot \mathbf{j}}{\tfrac{1}{2}\rho_{\infty}U^2_{\infty}A},
\]
where \(A\) is the projected frontal area of the sphere, and \(\mathbf{i}\) and \(\mathbf{j}\) are the unit vectors in the $x$ and $y$ directions, respectively. The Strouhal number is defined as 
\[
St=\frac{fD}{U_{\infty}},
\] 
where $f$ is the shedding frequency obtained by spectral analysis of the lift coefficient history. Table \ref{table:1} summarizes the comparison of the mean drag coefficient $\overline{C}_d$ and Strouhal number $St$ with values reported in the literature. Order $N=3$ and $N=4$ results shows a very small percentage of difference, $0.8\%$, for the mean drag coefficient and good agreement on the Strouhal number. The mean drag coefficients obtained with both approximation orders lie between the values reported in previous DNS \cite{rodriguez2011direct} and LES \cite{yun2006vortical} studies for the same configuration. The discrepancy with both references is approximately 5\%. A closer agreement is observed with the recent work of \cite{zhang2022dynamics}, who investigated the same problem using the high-order spectral element solver \texttt{Nek5000} \cite{fischer2015nek5000} by means of DNS. The vortex-shedding frequency, expressed by the Strouhal number $St$, also falls within the range of experimental and numerical results reported in the literature.

\begin{figure}[htbp!]
    \centering
        \begin{subfigure}[b]{0.45\textwidth}
            \includegraphics[width=\textwidth]{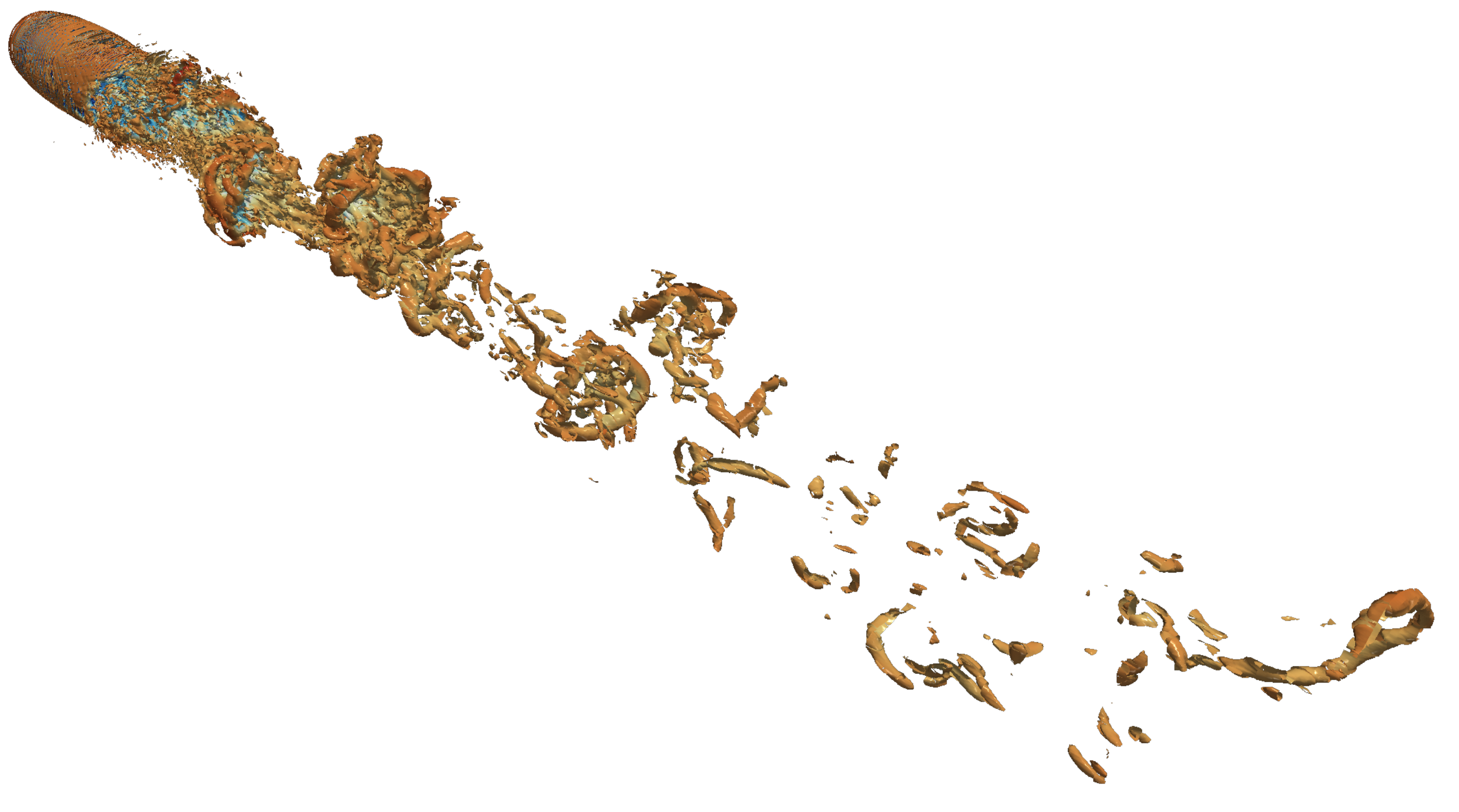}
            \caption{$t^*=50$}
        \end{subfigure}
        \begin{subfigure}[b]{0.45\textwidth}
            \includegraphics[width=\textwidth]{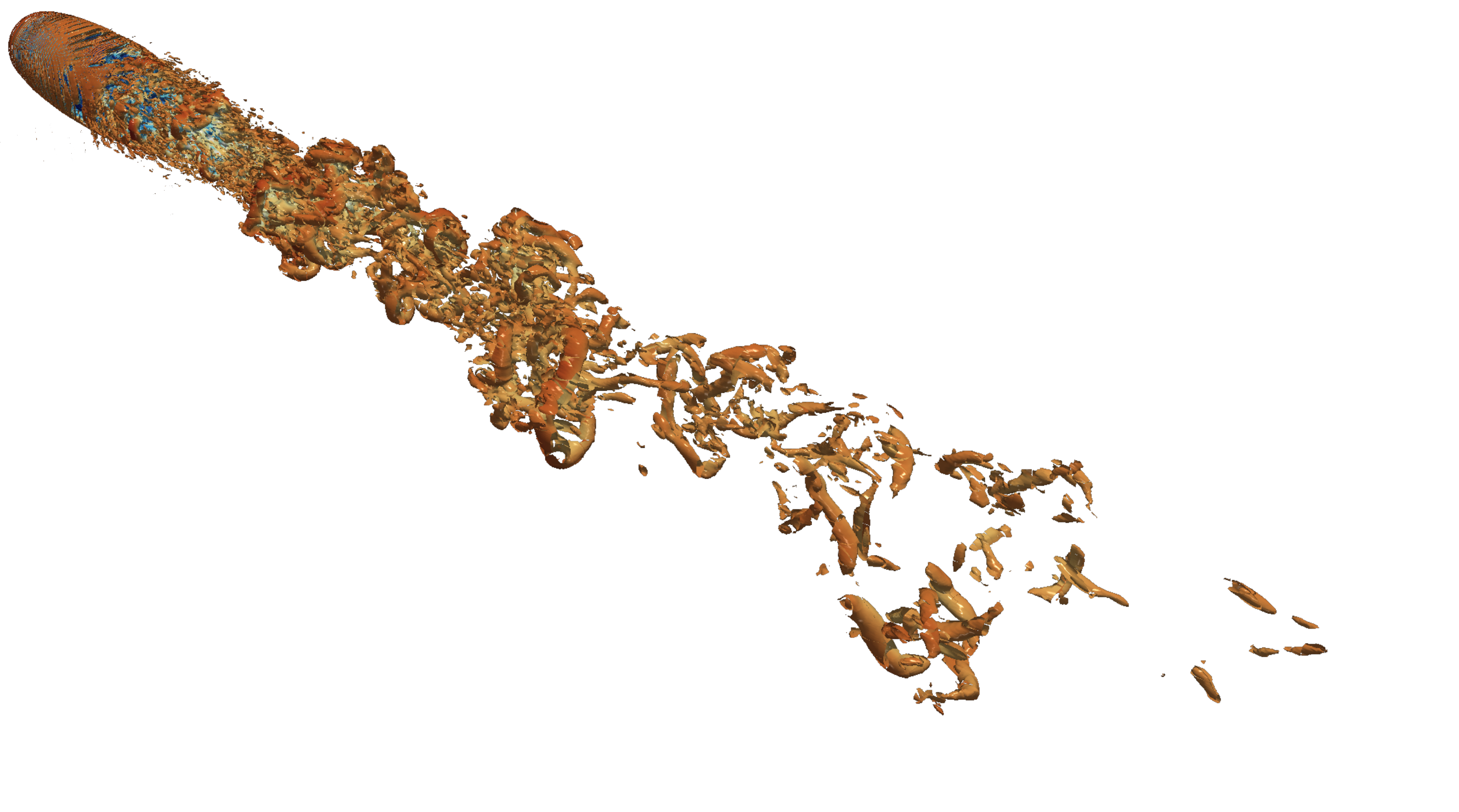}
            \caption{$t^*=90$}
        \end{subfigure}
        \begin{subfigure}[b]{0.45\textwidth}
           \includegraphics[width=\textwidth]{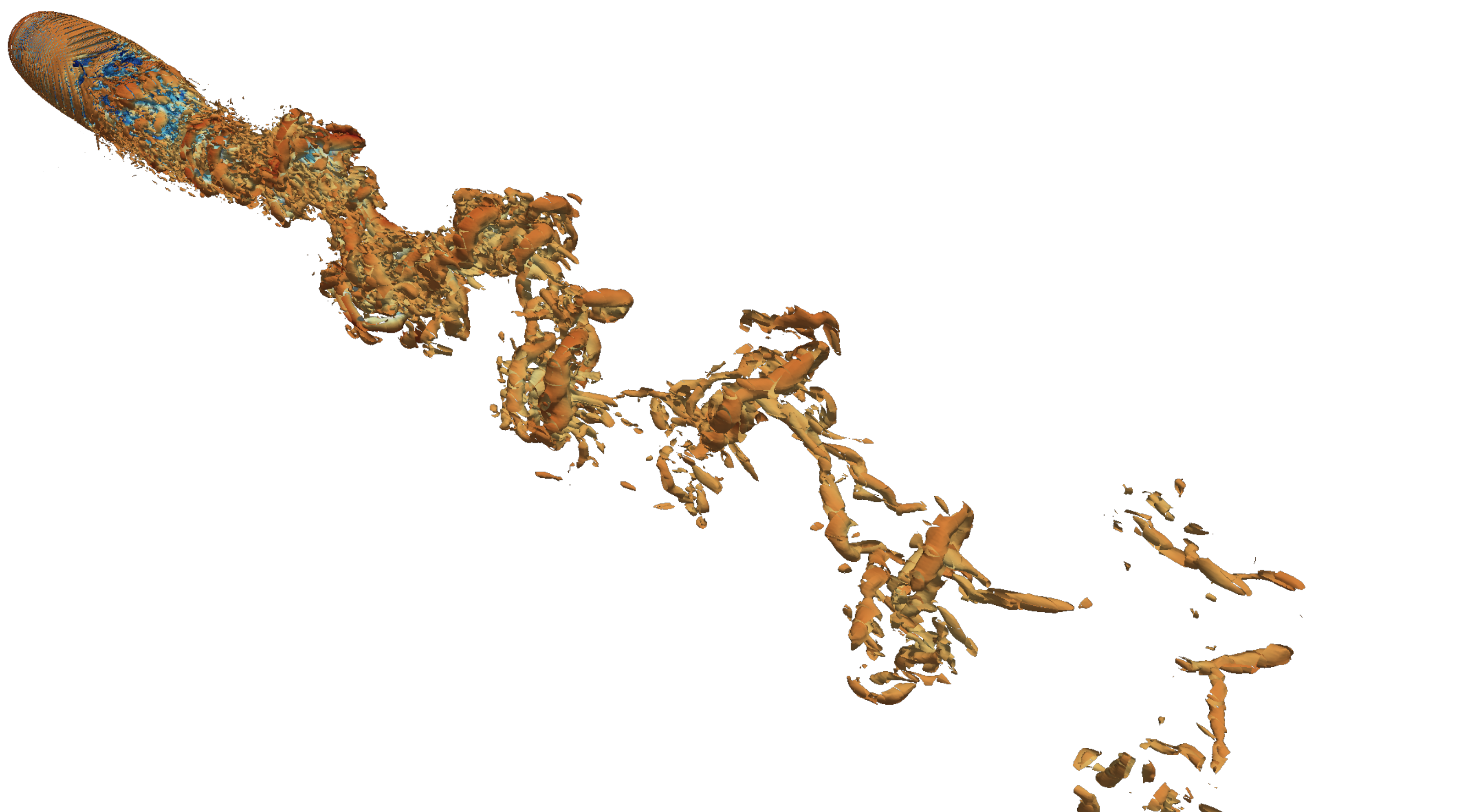}
           \caption{$t^*=120$}
        \end{subfigure}
\caption{Iso-surfaces of the Q-criterion, $Q=0.005$, colored by velocity magnitude, from the solution of the flow over sphere problem.}
    \label{fig:Q_Crit_sphere}
\end{figure}

\begin{figure}[htbp!]
    \centering
        \begin{subfigure}[b]{0.6\textwidth}
            \includegraphics[width=\textwidth]{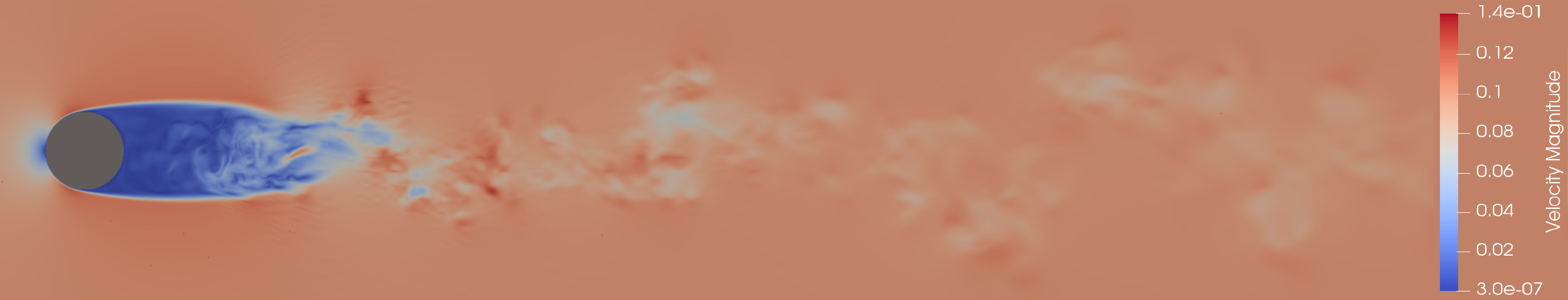}
            \caption{$t^*=50$}
        \end{subfigure} \\
        \begin{subfigure}[b]{0.6\textwidth}
            \includegraphics[width=\textwidth]{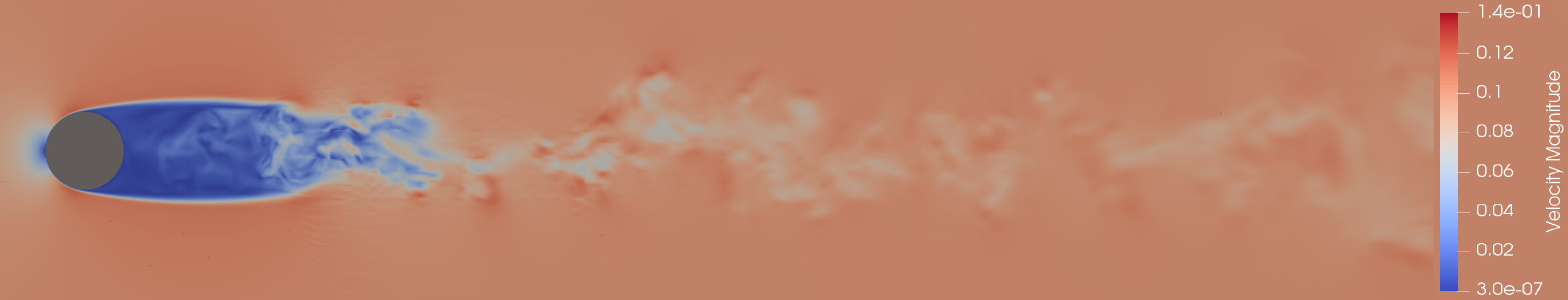}
            \caption{$t^*=90$}
        \end{subfigure}
        \begin{subfigure}[b]{0.6\textwidth}
            \includegraphics[width=\textwidth]
            {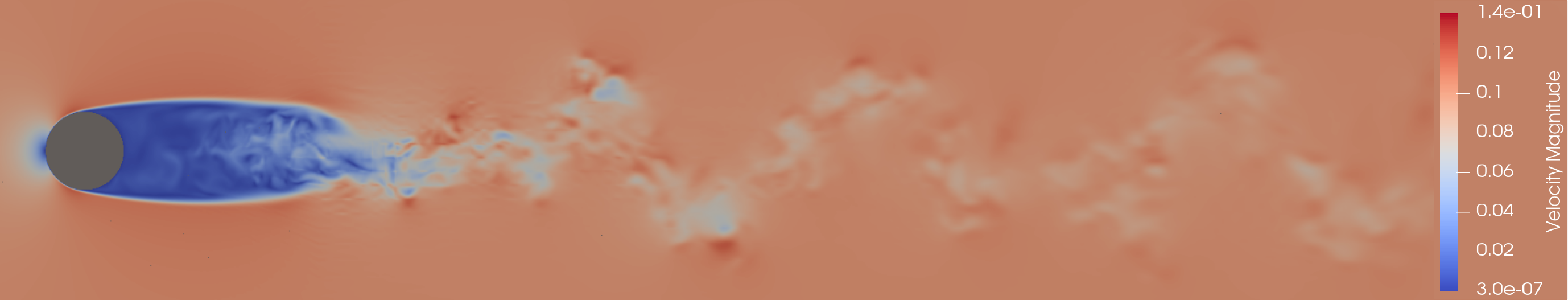}
            \caption{$t^*=120$}
        \end{subfigure}
        %\begin{subfigure}[b]{0.6\textwidth}
          % \includegraphics[width=\textwidth]{figures/V_t150_n3_newmesh.pdf}
        %\end{subfigure}
\caption{Velocity magnitude contours on $xy-$plane from the solution of the flow over sphere problem.}
    \label{fig:V_Mag_sphere}
\end{figure}

Figure \ref{fig:Q_Crit_sphere} illustrates the formation of vortical structures around the sphere after the transition to turbulence obtained with $N=4$. These structures develop into a helical-like shedding pattern with large-scale waviness as time progresses, consistent with the observations of \cite{yun2006vortical,rodriguez2011direct}. For the same time instants, the velocity magnitude contours around the sphere are shown in Figure \ref{fig:V_Mag_sphere}. The velocity profiles indicate that the recirculation bubble behind the sphere extends to approximately $x^* \approx 2D$, in good agreement with the results reported in the literature \cite{yun2006vortical,rodriguez2011direct}.

% \begin{table}[h!]
% \centering
% \begin{tabular}{|c c c c|} 
%  \hline
%   & Re & $S_t$ & $\overline{C_d}$ \\ [0.5ex] 
%  \hline\hline
%  Present Study (iLES, N=3) & 3700 & 0.219 &  0.369  \\ 
%  \hline
%   Present Study (iLES, N=4) & 3700 & - & -  \\ 
%  \hline
%  Rodriguez $et \ al.$ (DNS) \cite{rodriguez2011direct} & 3700 & 0.215 & 0.394 \\
%  \hline
%  Yun $et \ al.$ (LES) \cite{yun2006vortical} & 3700 & 0.21 & 0.355 \\
%  \hline
%  Kim $\&$ Durbin (exp.) \cite{kim1988observations}   & 3700 & 0.225 & $-$ \\
%  \hline
%  Sakamoto $\&$ Haniu (exp.) \cite{sakamoto1990study} & 3700 & 0.204 & $-$ \\ [1ex] 
%  \hline
% \end{tabular}
% \caption{Statistical flow properties for flow over sphere problem. Results are compared with experimental and numerical results from the literature.}
% \label{table:1}
% \end{table}

\begin{table}[h!]
\centering
\begin{tabular}{ c c c c } 
 \hline \hline
  & Re & $S_t$ & $\overline{C_d}$ \\ [0.5ex] 
 \hline\hline
 Present Study (N=3) & 3700 & 0.219 &  0.369  \\ 
 % \hline
 Present Study (N=4) & 3700 & 0.220 & 0.372  \\ 
  \hline
 Zhang and Peet (DNS) \cite{zhang2022dynamics} & 3700 & 0.220 & 0.375 \\
 Rodriguez $et \ al.$ (DNS) \cite{rodriguez2011direct} & 3700 & 0.215 & 0.394 \\
 % \hline
 Yun $et \ al.$ (LES) \cite{yun2006vortical} & 3700 & 0.21 & 0.355 \\

 \hline
 Kim $\&$ Durbin (exp.) \cite{kim1988observations}   & 3700 & 0.225 & $-$ \\
 % \hline
 Sakamoto $\&$ Haniu (exp.) \cite{sakamoto1990study} & 3700 & 0.204 & $-$ \\ [1ex] 
 \hline \hline
\end{tabular}
\caption{Statistical flow properties for flow over sphere problem. Results are compared with experimental and numerical results from the literature.}
\label{table:1}
\end{table}

\section{Conclusion}
\label{sec:conclusion}

Conclusively, we demonstrated the capability of a high-order DG formulation of the continuous Boltzmann equation to capture complex, turbulent flow physics for nearly incompressible flows. Our methodology relies on exploiting the dissipation mechanism of the numerical discretization to stably filter out the smallest scales. The relatively simple structure of the resulting equation system makes the method well-suited for controlling numerical dissipation. To prevent aliasing errors, we applied over-integration to the only nonlinear term in the semi-discrete system, namely the collision term. The linear advection term and the associated numerical flux were formulated with two different flux schemes in order to assess their effects on numerical dissipation and the stability of the method.

First, we studied the Taylor–Green vortex problem and analyzed nearly all aspects of the solution. We investigated the dissipation mechanisms by considering different orders of approximation, numbers of element and also two different numerical flux formulations. We observed that the LLF flux method is more dissipative than the upwind method. The discrepancy is especially more considerable for coarse mesh and lower orders. Higher polynomial orders and finer grids effectively suppress excessive dissipation introduced by more diffusive LLF flux formulation. The results obtained for first order properties and the energy spectra suggests that numerical dissipation is small and conserved in the high wave number regime that is one of the requirements of LES. DG–Boltzmann formulation accurately reproduces the decay of kinetic energy and the evolution of turbulence across a wide range of spatial resolutions and polynomial orders. 

Second, the method was applied to the flow over a sphere at a subcritical Reynolds number, where laminar–turbulent transition occurs with a large separation bubble. The complex transitional wake dynamics were accurately resolved and represented by Q-criterion and velocity contours demonstrating that the proposed scheme possesses sufficient ILES properties. The mean aerodynamic quantities were found to be in good quantitative agreement with reference numerical and experimental data.

Overall, this work highlights the potential of the DG–Boltzmann formulation as a high-order framework for ILES of turbulent flows. Future research will focus on incorporating adaptive strategies to enhance computational efficiency and better capture multiscale flow features in more complex turbulent configurations.

\section{Acknowledgements}
This research used resources of the Argonne Leadership Computing Facility, a U.S. Department of Energy (DOE) Office of Science user facility at Argonne National Laboratory and is based on research supported by the U.S. DOE Office of Science-Advanced Scientific Computing Research Program, under Contract No. DE-AC02-06CH11357.

The numerical calculations reported in this paper were partially performed at TUBITAK ULAKBIM, High Performance and Grid Computing Center (TRUBA resources).

\bibliographystyle{ieeetr}
\bibliography{main}

@article{park2017high,
  title={High-order implicit large-eddy simulations of flow over a {NACA0021} aerofoil},
  author={Park, JS and Witherden, FD and Vincent, PE},
  journal={AIAA Journal},
  volume={55},
  number={7},
  pages={2186--2197},
  year={2017},
  publisher={American Institute of Aeronautics and Astronautics}
}

@article{matsuyama2023implicit,
  title={Implicit large-eddy simulation of turbulent plane jet at {Re}= 104},
  author={Matsuyama, Shingo},
  journal={Computers \& Fluids},
  volume={250},
  pages={105732},
  year={2023},
  publisher={Elsevier}
}

@article{vermeire2016implicit,
  title={Implicit large eddy simulation using the high-order correction procedure via reconstruction scheme},
  author={Vermeire, Brian C and Nadarajah, Siva and Tucker, Paul G},
  journal={International Journal for Numerical Methods in Fluids},
  volume={82},
  number={5},
  pages={231--260},
  year={2016},
  publisher={Wiley Online Library}
}

@article{pope2001turbulent,
  title={Turbulent flows},
  author={Pope, Stephen B},
  journal={Measurement Science and Technology},
  volume={12},
  number={11},
  pages={2020--2021},
  year={2001}
}

@article{orszag1972numerical,
  title={Numerical simulation of three-dimensional homogeneous isotropic turbulence},
  author={Orszag, Steven A and Patterson Jr, GS},
  journal={Physical review letters},
  volume={28},
  number={2},
  pages={76},
  year={1972},
  publisher={APS}
}

@book{rogallo1981numerical,
  title={Numerical experiments in homogeneous turbulence},
  author={Rogallo, Robert Sugden},
  volume={81315},
  year={1981},
  publisher={National Aeronautics and Space Administration}
}

@article{moin1998direct,
  title={Direct numerical simulation: a tool in turbulence research},
  author={Moin, Parviz and Mahesh, Krishnan},
  journal={Annual review of fluid mechanics},
  volume={30},
  number={1},
  pages={539--578},
  year={1998},
  publisher={Annual Reviews 4139 El Camino Way, PO Box 10139, Palo Alto, CA 94303-0139, USA}
}

@article{drikakis2009large,
  title={Large eddy simulation using high-resolution and high-order methods},
  author={Drikakis, Dimitris and Hahn, Marco and Mosedale, Andrew and Thornber, Ben},
  journal={Philosophical Transactions of the Royal Society A: Mathematical, Physical and Engineering Sciences},
  volume={367},
  number={1899},
  pages={2985--2997},
  year={2009},
  publisher={The Royal Society London}
}

@article{sagaut2009large,
  title={Large eddy simulation for aerodynamics: status and perspectives},
  author={Sagaut, Pierre and Deck, S{\'e}bastien},
  journal={Philosophical Transactions of the Royal Society A: Mathematical, Physical and Engineering Sciences},
  volume={367},
  number={1899},
  pages={2849--2860},
  year={2009},
  publisher={The Royal Society London}
}

@inproceedings{boris2005large,
  title={On large eddy simulation using subgrid turbulence models comment 1},
  author={Boris, Jay P},
  booktitle={Whither Turbulence? Turbulence at the Crossroads: Proceedings of a Workshop Held at Cornell University, Ithaca, NY, March 22--24, 1989},
  pages={344--353},
  year={2005},
  organization={Springer}
}

@article{margolin2005design,
  title={The design and construction of implicit {LES} models},
  author={Margolin, LG and Rider, WJ},
  journal={International Journal for Numerical Methods in Fluids},
  volume={47},
  number={10-11},
  pages={1173--1179},
  year={2005},
  publisher={Wiley Online Library}
}

@article{margolin2006modeling,
  title={Modeling turbulent flow with implicit {LES}},
  author={Margolin, Len G and Rider, William J and Grinstein, Fernando F},
  journal={Journal of Turbulence},
  number={7},
  pages={N15},
  year={2006},
  publisher={Taylor \& Francis}
}

@article{ferrer2017interior,
  title={An interior penalty stabilised incompressible discontinuous {Galerkin}--{Fourier} solver for implicit large eddy simulations},
  author={Ferrer, Esteban},
  journal={Journal of Computational Physics},
  volume={348},
  pages={754--775},
  year={2017},
  publisher={Elsevier}
}

@article{parsani2010implicit,
  title={An implicit high-order spectral difference approach for large eddy simulation},
  author={Parsani, Matteo and Ghorbaniasl, Ghader and Lacor, Chris and Turkel, Eli},
  journal={Journal of Computational Physics},
  volume={229},
  number={14},
  pages={5373--5393},
  year={2010},
  publisher={Elsevier}
}

@article{kopriva1996conservative,
  title={A conservative staggered-grid {Chebyshev} multidomain method for compressible flows},
  author={Kopriva, David A and Kolias, John H},
  journal={Journal of Computational Physics},
  volume={125},
  number={1},
  pages={244--261},
  year={1996},
  publisher={Elsevier}
}

@article{liu2006spectral,
  title={Spectral difference method for unstructured grids I: Basic formulation},
  author={Liu, Yen and Vinokur, Marcel and Wang, Zhi Jian},
  journal={Journal of Computational Physics},
  volume={216},
  number={2},
  pages={780--801},
  year={2006},
  publisher={Elsevier}
}

@article{wang2007spectral,
  title={Spectral difference method for unstructured grids II: extension to the {Euler} equations},
  author={Wang, Zhi Jian and Liu, Yen and May, Georg and Jameson, Antony},
  journal={Journal of Scientific Computing},
  volume={32},
  pages={45--71},
  year={2007},
  publisher={Springer}
}

@book{cockburn2012discontinuous,
  title={Discontinuous {Galerkin} methods: theory, computation and applications},
  author={Cockburn, Bernardo and Karniadakis, George E and Shu, Chi-Wang},
  volume={11},
  year={2012},
  publisher={Springer Science \& Business Media}
}

@inproceedings{huynh2007flux,
  title={A flux reconstruction approach to high-order schemes including discontinuous {Galerkin} methods},
  author={Huynh, Hung T},
  booktitle={18th AIAA Computational Fluid Dynamics Conference},
  pages={4079},
  year={2007}
}

@book{hesthaven2007nodal,
  title={Nodal discontinuous {Galerkin} methods: algorithms, analysis, and applications},
  author={Hesthaven, Jan S and Warburton, Tim},
  year={2007},
  publisher={Springer Science \& Business Media}
}

@article{gassner2013accuracy,
  title={On the accuracy of high-order discretizations for underresolved turbulence simulations},
  author={Gassner, Gregor J and Beck, Andrea D},
  journal={Theoretical and Computational Fluid Dynamics},
  volume={27},
  number={3},
  pages={221--237},
  year={2013},
  publisher={Springer}
}

@article{beck2014high,
  title={High-order discontinuous {Galerkin} spectral element methods for transitional and turbulent flow simulations},
  author={Beck, Andrea D and Bolemann, Thomas and Flad, David and Frank, Hannes and Gassner, Gregor J and Hindenlang, Florian and Munz, Claus-Dieter},
  journal={International Journal for Numerical Methods in Fluids},
  volume={76},
  number={8},
  pages={522--548},
  year={2014},
  publisher={Wiley Online Library}
}

@article{bhatnagar1954model,
  title={A model for collision processes in gases. {I.} {Small} amplitude processes in charged and neutral one-component systems},
  author={Bhatnagar, Prabhu Lal and Gross, Eugene P and Krook, Max},
  journal={Physical Review},
  volume={94},
  number={3},
  pages={511},
  year={1954},
  publisher={APS}
}

@article{taylor1937mechanism,
  title={Mechanism of the production of small eddies from large ones},
  author={Taylor, Geoffrey Ingram and Green, Albert Edward},
  journal={Proceedings of the Royal Society of London. Series A-Mathematical and Physical Sciences},
  volume={158},
  number={895},
  pages={499--521},
  year={1937},
  publisher={The Royal Society London}
}

@article{carton2014assessment,
  title={Assessment of a discontinuous {Galerkin} method for the simulation of vortical flows at high {Reynolds} number},
  author={Carton de Wiart, Corentin and Hillewaert, Koen and Duponcheel, Matthieu and Winckelmans, Gr{\'e}goire},
  journal={International Journal for Numerical Methods in Fluids},
  volume={74},
  number={7},
  pages={469--493},
  year={2014},
  publisher={Wiley Online Library}
}

@article{Mengaldo2018,
   author = {G. Mengaldo and R. C. Moura and B. Giralda and J. Peiró and S. J. Sherwin},
   doi = {10.1016/j.compfluid.2017.09.016},
   issn = {00457930},
   journal = {Computers and Fluids},
   month = {6},
   pages = {349-364},
   publisher = {Elsevier Ltd},
   title = {Spatial eigensolution analysis of discontinuous {Galerkin} schemes with practical insights for under-resolved computations and implicit {LES}},
   volume = {169},
   year = {2018},
}

@article{Uranga2011,
   author = {A. Uranga and P. O. Persson and M. Drela and J. Peraire},
   doi = {10.1002/nme.3036},
   issn = {00295981},
   issue = {1-5},
   journal = {International Journal for Numerical Methods in Engineering},
   month = {7},
   pages = {232-261},
   title = {Implicit Large Eddy Simulation of transition to turbulence at low {Reynolds} numbers using a Discontinuous {Galerkin} method},
   volume = {87},
   year = {2011},
}

@article{moura2017eddy,
  title={On the eddy-resolving capability of high-order discontinuous {Galerkin} approaches to implicit {LES/under-resolved} {DNS} of {Euler} turbulence},
  author={Moura, Rodrigo Costa and Mengaldo, Gianmarco and Peir{\'o}, Joaquim and Sherwin, Spencer J},
  journal={Journal of Computational Physics},
  volume={330},
  pages={615--623},
  year={2017},
  publisher={Elsevier}
}

@article{Moura2017,
   author = {R. C. Moura and G. Mengaldo and J. Peiró and S. J. Sherwin},
   doi = {10.1016/j.jcp.2016.10.056},
   issn = {10902716},
   journal = {Journal of Computational Physics},
   month = {2},
   pages = {615-623},
   publisher = {Academic Press Inc.},
   title = {On the eddy-resolving capability of high-order discontinuous {Galerkin} approaches to implicit {LES} / under-resolved {DNS} of {Euler} turbulence},
   volume = {330},
   year = {2017},
}

@article{Fernandez2017,
   author = {P. Fernandez and N. C. Nguyen and J. Peraire},
   doi = {10.1016/j.jcp.2017.02.015},
   issn = {10902716},
   journal = {Journal of Computational Physics},
   month = {5},
   pages = {308-329},
   publisher = {Academic Press Inc.},
   title = {The hybridized Discontinuous {Galerkin} method for Implicit Large-Eddy Simulation of transitional turbulent flows},
   volume = {336},
   year = {2017},
}

@article{Karakus2019,
   author = {A. Karakus and N. Chalmers and J. S. Hesthaven and T. Warburton},
   doi = {10.1016/j.jcp.2019.03.050},
   issn = {10902716},
   journal = {Journal of Computational Physics},
   month = {8},
   pages = {175-202},
   publisher = {Academic Press Inc.},
   title = {Discontinuous {Galerkin} discretizations of the {Boltzmann–BGK} equations for nearly incompressible flows: {Semi-analytic} time stepping and absorbing boundary layers},
   volume = {390},
   year = {2019},
}

@article{krank2017high,
  title={A high-order semi-explicit discontinuous {Galerkin} solver for {3D} incompressible flow with application to {DNS} and {LES} of turbulent channel flow},
  author={Krank, Benjamin and Fehn, Niklas and Wall, Wolfgang A and Kronbichler, Martin},
  journal={Journal of Computational Physics},
  volume={348},
  pages={634--659},
  year={2017},
  publisher={Elsevier}
}

@article{lederer2023high,
  title={High-order projection-based upwind method for implicit large eddy simulation},
  author={Lederer, Philip L and Mooslechner, Xaver and Sch{\"o}berl, Joachim},
  journal={Journal of Computational Physics},
  volume={493},
  pages={112492},
  year={2023},
  publisher={Elsevier}
}

@inproceedings{debonis2013solutions,
  title={Solutions of the {Taylor-Green} vortex problem using high-resolution explicit finite difference methods},
  author={DeBonis, James},
  booktitle={51st AIAA Aerospace Sciences Meeting Including the New Horizons Forum and Aerospace Exposition},
  pages={382},
  year={2013}
}

@article{van2011comparison,
  title={A comparison of vortex and pseudo-spectral methods for the simulation of periodic vortical flows at high {Reynolds} numbers},
  author={Van Rees, Wim M and Leonard, Anthony and Pullin, Dale I and Koumoutsakos, Petros},
  journal={Journal of Computational Physics},
  volume={230},
  number={8},
  pages={2794--2805},
  year={2011},
  publisher={Elsevier}
}

@article{brachet1983small,
  title={Small-scale structure of the {Taylor--Green} vortex},
  author={Brachet, Marc E and Meiron, Daniel I and Orszag, Steven A and Nickel, BG and Morf, Rudolf H and Frisch, Uriel},
  journal={Journal of Fluid Mechanics},
  volume={130},
  pages={411--452},
  year={1983},
  publisher={Cambridge University Press}
}

@article{zhu2016implicit,
  title={Implicit large-eddy simulation for the high-order flux reconstruction method},
  author={Zhu, Hui and Fu, Song and Shi, Lei and Wang, ZJ},
  journal={AIAA Journal},
  volume={54},
  number={9},
  pages={2721--2733},
  year={2016},
  publisher={American Institute of Aeronautics and Astronautics}
}

@article{tolke_discretization_2000,
	title = {Discretization of the {Boltzmann} equation in velocity space using a {Galerkin} approach},
	volume = {129},
	issn = {0010-4655},
	url = {https://www.sciencedirect.com/science/article/pii/S0010465500000965},
	doi = {10.1016/S0010-4655(00)00096-5},
	number = {1},
	urldate = {2024-04-25},
	journal = {Computer Physics Communications},
	author = {Tölke, Jonas and Krafczyk, Manfred and Schulz, Manuel and Rank, Ernst},
	year = {2000},
	pages = {91--99},
}

@article{bull2015simulation,
  title={Simulation of the {Taylor--Green} vortex using high-order flux reconstruction schemes},
  author={Bull, Jonathan R and Jameson, Antony},
  journal={AIAA Journal},
  volume={53},
  number={9},
  pages={2750--2761},
  year={2015},
  publisher={American Institute of Aeronautics and Astronautics}
}

@article{hamedi2022optimized,
  title={Optimized filters for stabilizing high-order large eddy simulation},
  author={Hamedi, Mohsen and Vermeire, Brian C},
  journal={Computers \& Fluids},
  volume={237},
  pages={105301},
  year={2022},
  publisher={Elsevier}
}

@MISC{ChalmersKarakusAustinSwirydowiczWarburton2020,
      author = "Chalmers, N. and Karakus, A. and Austin, A. P. and Swirydowicz, K. and Warburton, T.",
      title = "{libParanumal}: a performance portable high-order finite element library",
      year = "2022",
      url = "https://github.com/paranumal/libparanumal",
      doi = "10.5281/zenodo.4004744",
      note = "Release 0.5.0"}

@article{rodriguez2011direct,
  title={Direct numerical simulation of the flow over a sphere at {Re= 3700}},
  author={Rodriguez, Ivette and Borell, Ricard and Lehmkuhl, Oriol and Segarra, Carlos D Perez and Oliva, Assensi},
  journal={Journal of Fluid Mechanics},
  volume={679},
  pages={263--287},
  year={2011},
  publisher={Cambridge University Press}
}

@article{kim1988observations,
  title={Observations of the frequencies in a sphere wake and of drag increase by acoustic excitation},
  author={Kim, HJ and Durbin, PA},
  journal={The Physics of fluids},
  volume={31},
  number={11},
  pages={3260--3265},
  year={1988},
  publisher={American Institute of Physics}
}

@article{sakamoto1990study,
    author = {Sakamoto, H. and Haniu, H.},
    title = {A Study on Vortex Shedding From Spheres in a Uniform Flow},
    journal = {Journal of Fluids Engineering},
    volume = {112},
    number = {4},
    pages = {386-392},
    year = {1990},
    issn = {0098-2202},
    doi = {10.1115/1.2909415},
    url = {https://doi.org/10.1115/1.2909415},
}

@article{yun2006vortical,
  title={Vortical structures behind a sphere at subcritical {Reynolds} numbers},
  author={Yun, Giwoong and Kim, Dongjoo and Choi, Haecheon},
  journal={Physics of Fluids},
  volume={18},
  number={1},
  year={2006},
  publisher={AIP Publishing}
}

@article{hunt1988eddies,
  title={Eddies, streams, and convergence zones in turbulent flows},
  author={Hunt, Julian CR and Wray, Alan A and Moin, Parviz},
  journal={Studying turbulence using numerical simulation databases, 2. Proceedings of the 1988 summer program},
  year={1988}
}

@article{shi2003discontinuous,
  title={ {Discontinuous} {Galerkin} spectral element lattice {Boltzmann method} on triangular element},
  author={Shi, Xing and Lin, Jianzhong and Yu, Zhaosheng},
  journal={International Journal for Numerical Methods in Fluids},
  volume={42},
  number={11},
  pages={1249--1261},
  year={2003},
  publisher={Wiley Online Library}
}

@article{duster2006high,
  title={High-order finite elements applied to the discrete {Boltzmann} equation},
  author={D{\"u}ster, Alexander and Demkowicz, Leszek and Rank, Ernst},
  journal={International Journal for Numerical Methods in Engineering},
  volume={67},
  number={8},
  pages={1094--1121},
  year={2006},
  publisher={Wiley Online Library}
}

@article{min2011spectral,
  title={A spectral-element discontinuous {Galerkin} lattice {Boltzmann} method for nearly incompressible flows},
  author={Min, Misun and Lee, Taehun},
  journal={Journal of Computational Physics},
  volume={230},
  number={1},
  pages={245--259},
  year={2011},
  publisher={Elsevier}
}

@article{fehn2019high,
  title={High-order {DG} solvers for underresolved turbulent incompressible flows: {A} comparison of {L2} and {H}(div) methods},
  author={Fehn, Niklas and Kronbichler, Martin and Lehrenfeld, Christoph and Lube, Gert and Schroeder, Philipp W},
  journal={International Journal for Numerical Methods in Fluids},
  volume={91},
  number={11},
  pages={533--556},
  year={2019},
  publisher={Wiley Online Library}
}

@article{winters2018comparative,
  title={A comparative study on polynomial dealiasing and split form discontinuous {Galerkin} schemes for under-resolved turbulence computations},
  author={Winters, Andrew R and Moura, Rodrigo C and Mengaldo, Gianmarco and Gassner, Gregor J and Walch, Stefanie and Peiro, Joaquim and Sherwin, Spencer J},
  journal={Journal of Computational Physics},
  volume={372},
  pages={1--21},
  year={2018},
  publisher={Elsevier}
}

@article{moura2024joint,
  title={Joint-mode diffusion analysis of discontinuous {Galerkin} methods: {Towards} superior dissipation estimates for nonlinear problems and implicit {LES}},
  author={Moura, Rodrigo Costa and Fernandes, Lucas Dantas and da Silva, AFC and Sherwin, Spencer J},
  journal={Journal of Computational Physics},
  volume={505},
  pages={112912},
  year={2024},
  publisher={Elsevier}
}

@article{de2018use,
  title={On the use of a high-order discontinuous {Galerkin} method for {DNS} and {LES} of wall-bounded turbulence},
  author={de la Llave Plata, Marta and Couaillier, Vincent and Le Pape, Marie-Claire},
  journal={Computers \& Fluids},
  volume={176},
  pages={320--337},
  year={2018},
  publisher={Elsevier}
}

@incollection{bolemann2015high,
  title={High-order discontinuous {Galerkin} schemes for large-eddy simulations of moderate {Reynolds} number flows},
  author={Bolemann, T and Beck, A and Flad, D and Frank, H and Mayer, V and Munz, C-D},
  booktitle={IDIHOM: Industrialization of High-Order Methods-A Top-Down Approach: Results of a Collaborative Research Project Funded by the European Union, 2010-2014},
  pages={435--456},
  year={2015},
  publisher={Springer}
}

@article{wang2002spectral,
	title = {Spectral ({Finite}) {Volume} {Method} for {Conservation} {Laws} on {Unstructured} {Grids}. {Basic} {Formulation}: {Basic} {Formulation}},
	volume = {178},
	issn = {0021-9991},
	shorttitle = {Spectral ({Finite}) {Volume} {Method} for {Conservation} {Laws} on {Unstructured} {Grids}. {Basic} {Formulation}},
	url = {https://www.sciencedirect.com/science/article/pii/S0021999102970415},
	doi = {10.1006/jcph.2002.7041},
	number = {1},
	urldate = {2025-08-16},
	journal = {Journal of Computational Physics},
	author = {Wang, Z. J.},
	month = may,
	year = {2002},
	pages = {210--251},
}

@article{wang2021implicit,
  title={Implicit large-eddy simulations of turbulent flow in a channel via spectral/hp element methods},
  author={Wang, Rui and Wu, Feng and Xu, Hui and Sherwin, Spencer J},
  journal={Physics of Fluids},
  volume={33},
  number={3},
  year={2021},
  publisher={AIP Publishing}
}

@inproceedings{zhou_implicit_2010,
	address = {Chicago, Illinois},
	title = {Implicit {Large} {Eddy} {Simulation} of {Low} {Reynolds} {Number} {Transitional} {Flow} over a {Wing} {Using} {High}-{Order} {Spectral} {Difference} {Method}},
	isbn = {9781600869563},
	url = {https://arc.aiaa.org/doi/10.2514/6.2010-4442},
	doi = {10.2514/6.2010-4442},
	language = {en},
	urldate = {2025-08-16},
	booktitle = {40th {Fluid} {Dynamics} {Conference} and {Exhibit}},
	publisher = {American Institute of Aeronautics and Astronautics},
	author = {Zhou, Ying and Wang, Zhi J},
	month = jun,
	year = {2010},
}

@inproceedings{zhang2022dynamics,
  title={The dynamics of coherent structures in a turbulent wake past a sphere at {Re= 3700}},
  author={Zhang, Fengrui and Peet, Yulia T},
  booktitle={Turbulence and shear flow phenomena},
  volume={2022},
  year={2022},
  organization={Twelfth International Symposium on Turbulence and Shear Flow Phenomena}
}

@article{fischer2015nek5000,
  title={Nek5000: {User’s} manual},
  author={Fischer, Paul and Lottes, James and Kerkemeier, Stefan and Marin, Oana and Heisey, Katherine and Obabko, Aleks and Merzari, Elia and Peet, Yulia},
  journal={Argonne National Laboratory, Lemont, IL, Technical Report No. ANL/MCS-TM-351},
  year={2015}
}

@article{chan2017penalty,
  title={On the penalty stabilization mechanism for upwind discontinuous {Galerkin} formulations of first order hyperbolic systems},
  author={Chan, Jesse and Warburton, T},
  journal={Computers \& Mathematics with Applications},
  volume={74},
  number={12},
  pages={3099--3110},
  year={2017},
  publisher={Elsevier}
}

@article{chan2017weight,
  title={Weight-adjusted discontinuous {Galerkin} methods: wave propagation in heterogeneous media},
  author={Chan, Jesse and Hewett, Russell J and Warburton, Timothy},
  journal={SIAM Journal on Scientific Computing},
  volume={39},
  number={6},
  pages={A2935--A2961},
  year={2017},
  publisher={SIAM}
}

@article{warburton2013low,
  title={A low-storage curvilinear discontinuous {Galerkin} method for wave problems},
  author={Warburton, Timothy},
  journal={SIAM Journal on Scientific Computing},
  volume={35},
  number={4},
  pages={A1987--A2012},
  year={2013},
  publisher={SIAM}
}

@article{kopriva2014energy,
  title={An energy stable discontinuous {Galerkin} spectral element discretization for variable coefficient advection problems},
  author={Kopriva, David A and Gassner, Gregor J},
  journal={SIAM Journal on Scientific Computing},
  volume={36},
  number={4},
  pages={A2076--A2099},
  year={2014},
  publisher={SIAM}
}

\end{document}